\documentclass[
reprint, amsfonts, 
amssymb, amsmath, aps, 
floatfix, twoside, 
superscriptaddress,
prx,
longbibliography,
]{revtex4-2}
\usepackage{graphicx}
\usepackage{dcolumn}
\usepackage[english]{babel}
\usepackage[utf8]{inputenc}
\usepackage{amsthm}
\usepackage{mathtools}
\usepackage{makecell}
\usepackage{physics}
\usepackage{xcolor,colortbl}
\usepackage{adjustbox}
\usepackage{placeins}
\usepackage[T1]{fontenc}
\usepackage{lipsum}
\usepackage{csquotes}
\usepackage{hyperref}
\usepackage[bb=boondox]{mathalfa}
\usepackage[normalem]{ulem}
\usepackage{diagbox}
\usepackage{slashbox}
\usepackage{soul}
\usepackage{multirow}

\newcommand{\ZPF}{\mathrm{ZPF}}

\begin{document}
\title{Characterization of drive-induced unwanted state transitions in superconducting circuits}

\makeatletter
\renewcommand*{\@fnsymbol}[1]{\ifcase#1\or i\or ii\or iii\or iv\or v\or vi\or vii\or viii\or ix\or x\else\@ctrerr\fi}
\makeatother
%%=============================================================
%% Author list
%%=============================================================
\author{W. Dai*}
\email{wei.dai.wd279@yale.edu}
\altaffiliation{Current address: Quantum Machines, USA}
\thanks{*Equal contributions.}
\author{S. Hazra*}
\email{sumeru.hazra@yale.edu}
\altaffiliation{Current address: Karlsruher Institut für Technologie, Germany}
\thanks{*Equal contributions.}
\author{D. K. Weiss*}
\email{daniel.weiss@yale.edu}
\altaffiliation{Current address: Quantum Circuits Inc., New Haven, Connecticut, USA}
\thanks{*Equal contributions.}
\author{P. D. Kurilovich}
\author{T. Connolly}
\author{H. K. Babla}
\author{S. Singh}
\altaffiliation{Current address: IBM Quantum, Yorktown Heights, New York, USA}
\author{V. R. Joshi}
\altaffiliation{Current address: Microsoft, Redmond, Washington, USA}
\author{A. Z. Ding}
\altaffiliation{Current address: AWS Center for Quantum Computing, Pasadena, CA, USA}
\author{P. D. Parakh}
\altaffiliation{Current address: Stanford University, California, USA}
\author{J. Venkatraman}
\altaffiliation{Current address: University of California, Santa Barbara, California, USA}
\author{X. Xiao}
\author{L. Frunzio}
\author{M. H. Devoret}
\email{michel.devoret@yale.edu}
\altaffiliation{Current address: Department of Physics, University of California, Santa Barbara, California 93106, USA and Google Quantum AI, Santa Barbara, California, USA}
\affiliation{Department of Physics and Applied Physics, Yale University, New Haven, Connecticut 06520, USA, and \\
Yale Quantum Institute, Yale University, New Haven, Connecticut 06520, USA}

%%=============================================================
%% Abstract
%%=============================================================
\date{\today} 

\begin{abstract}
Microwave drives are essential for implementing control and readout operations in superconducting quantum circuits. However, increasing the drive strength eventually leads to unwanted state transitions which limit the speed and fidelity of such operations.
In this work, we systematically investigate such transitions in a fixed-frequency qubit subjected to microwave drives spanning a 9 GHz frequency range. 
We identify the physical origins of these transitions and classify them into three categories. (1) Resonant energy exchange with parasitic two-level systems, activated by drive-induced ac-Stark shifts, (2) multi-photon transitions to non-computational states, intrinsic to the circuit Hamiltonian, 
and (3) inelastic scattering processes in which the drive causes a state transition in the superconducting circuit, while transferring excess energy to a spurious electromagnetic mode or two-level system (TLS) material defect. 
We show that the Floquet steady-state simulation, complemented by an electromagnetic simulation of the physical device, accurately predicts the observed transitions that do not involve TLS.
Our results provide a comprehensive classification of these transitions and offer mitigation strategies through informed choices of drive frequency as well as improved circuit design.
\end{abstract}

\maketitle
%%=============================================================
%% INTRODUCTION
%%=============================================================
% add line numbers
% \linenumbers

\section{Introduction}

Superconducting circuits have emerged as one of the leading platforms for quantum information processing, owing to their scalable architectures, flexible designs, and direct compatibility with microwave-based control and measurement systems~\cite{wallraff_2004, schuster_2007}. In this architecture, the Josephson junction provides the essential nonlinearity, while microwave drives underpin a broad range of operations, including qubit stabilization, manipulation, and measurement.
At the level of physical qubits, microwave drives~\cite{rigetti_2010, chow_2010, mit_flux_drive, floquet_fluxonium, Levine2024} are essential for implementing high-fidelity single- and two-qubit gates~\cite{li_zhiyuan_2023, ding_2023, li_rui_2024, rower_2024}, and for executing various readout protocols such as dispersive readout~\cite{Swiadek2023}, longitudinal readout~\cite{didier_2015}, conditional displacement readout~\cite{touzard_2019}, and cat-quadrature readout~\cite{grimm_2020}.  These control and readout operations form the basis of all discrete variable quantum error correction codes~\cite{krinner_2022, google_2023}.
Microwave drives also enable a broad set of capabilities for bosonic quantum error correction~\cite{gkp_2001, bosonic_2016}. They are central to cavity control~\cite{gao_2019, eickbusch_ecd_2022, chapman_bs_2023, ding2025}, the realization of error-corrected quantum memories~\cite{ofek_2016, binomial_2019, campagne_gkp_2020,  gertler_2021, Sivak2023_gkp, nord_2024, brock_2025}, and the stabilization of logical qubits using techniques such as reservoir engineering~\cite{leghtas_2015, lescanne_2020} and Hamiltonian engineering~\cite{grimm_2020, hajr_2024}. 

%%-------------------------------------------------------------
%% Figure 1: Motivation
%%-------------------------------------------------------------
\begin{figure*}[t]
\includegraphics[width = 0.96\textwidth]{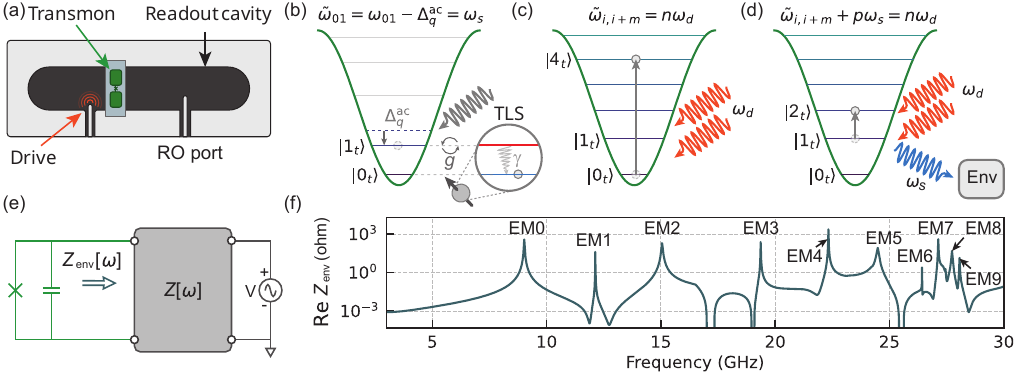}
\caption[Mechanisms of driven unwanted  state transitions in a transmon.]{Mechanisms of drive-induced unwanted  state transitions in a transmon. (a) A cartoon showing a transmon in a 3D cavity. The fundamental mode of the cavity plays the role of the readout resonator. Any parametric drive, including the readout pulse is delivered through designated ports coupled to the cavity.
The qubit is defined by the two lowest levels of the cosine potential of tranmson. 
(b) Mechanism \textit{A}: Resonant exchange of energy between the qubit and a spurious mode e.g. a two-level system (TLS). Under the drive, the transmon experiences ac-Stark shift and the qubit tunes into resonance with one or more TLS near the qubit frequency. Such an exchange depends only on the amount of ac-stark shift experienced by the qubit, and not on any frequency-matching condition involving the drive frequency. 
(c) Mechanism \textit{B}: Intrinsic multi-photon excitation of the transmon by absorbing multiple drive photons, due to its nonlinearity. When the energy gap between the initial and a higher excited state matches an integer multiple of the drive frequency (within selection rules), the transmon is excited through this multi-photon process. 
(d) Mechanism \textit{C}: Inelastic scattering of the drive off the transmon, causing it to either excite or relax while transferring the excess energy to the environment. (e) The transition rate of mechanism \textit{C} depends on the real part of the impedance seen by the transmon at the emission frequency\cite{connolly_2025}. 
In a realistic system, the transmon is coupled to a complex impedance network, $Z[\omega]$, whose details are often unknown to the experimentalist.  The peaks in ${\rm{Re}}~Z_{\rm{env}}$ corresponds to
additional modes arising from device geometry or material defects, that couple to the transmon.
(f) Simulated ${\rm{Re}}~Z[\omega]$ of the 3D transmon shown schematically in (a), showing multiple resonances due to spurious electromagnetic modes. Material defects introduce additional modes, not shown here, further crowding the spectrum. 
}
\label{fig:motivation} 
\end{figure*}
%%-------------------------------------------------------------

One of the limitations for the fidelity of such operations comes from the decoherence of the physical qubit, leading to incoherent errors. The probability of such errors can be reduced by increasing the speed of the operations, achieved by increasing the strength of the microwave drive. However, as the drive strength increases, the system experiences drive-induced unwanted state transitions (DUST). 
Such deleterious transitions can either happen between the computational states~\cite{Sivak2023_gkp,Thorbeck_2024_readout_T1,Bengtsson2024} or between the computational states and the non-computational ones~\cite{Sank_2016_MIST,Khezri_2023_MIST,Shillito2022,lu_2023, Cohen2023,Dumas_2024_Ionization,xia2025_threshold,Xiao2023,Nesterov_2024_MIST,Fechant2025, singh2024impact, bista2025readout}. The onset of errors caused by these transitions limits the drive strength the qubit can tolerate, and hence the maximum speed and fidelity one can achieve for quantum operations. DUST has been investigated in different superconducting circuit implementations, such as transmon~\cite{Sank_2016_MIST, Khezri_2023_MIST, Dumas_2024_Ionization, Thorbeck_2024_readout_T1}, fluxonium~\cite{bothara_2025, bista2025readout, singh2024impact,Nesterov_2024_MIST}, and driven stabilized cat qubits~\cite{grimm_2020, venkatraman_static_effective_2022, Xiao2023, Frattini2024, hajr_2024, ding2025}. However, in experiments, the understanding and mitigation of DUST are challenged by the existence of multiple competing mechanisms leading to similar observations. Can we parse the mechanisms of DUST observed in an experimental device and systematically avoid them in the design phase of the circuit?

In this work, we answer this question affirmatively by experimentally discriminating three distinct mechanisms responsible for DUST in a fixed-frequency 3D transmon: 

\textit{A. Qubit brought into resonance with a  spurious mode by ac-Stark shift.\textemdash} 
The transmon can resonantly exchange energy with a spurious mode, such as a TLS in the environment, leading to transmon decay, as shown in Fig.~\ref{fig:motivation}(b). 
This mechanism had previously been reported to be responsible for the reduction of the lifetime of driven superconducting qubits, ~\cite{Sivak2023_gkp,Thorbeck_2024_readout_T1,Bengtsson2024}. 

\textit{B. Intrinsic multi-photon excitation.\textemdash} Multiple drive photons are resonantly absorbed at once to excite the transmon to a non-computational state~\cite{Sank_2016_MIST, Khezri_2023_MIST, Xiao2023,Dumas_2024_Ionization,xia2025_threshold, Nesterov_2024_MIST}. For example, the transmon can absorb two drive photons and transition directly from the ground state to the fourth excited state, as shown in Fig.~\ref{fig:motivation}(c).

\textit{C. Inelastic scattering involving external modes.\textemdash} The drive causes a transmon transition and scatters into the transmon environment at a different frequency~\cite{singh2024impact}, as shown in Fig.~\ref{fig:motivation}(d). 
The rate of such a process is governed by the real part of the environmental impedance ${\rm{Re}}~Z_{\rm{env}}$ seen by the transmon [see Fig.~\ref{fig:motivation}(e)] at the frequency $\omega_s$ of the scattered photon~\cite{connolly_2025}. 
The environment external to the transmon contains several known and unknown degrees of freedom. 
Whenever such a mode is present in the transmon environment at the frequency $\omega_s$, the rate of the process is enhanced. For example, the electromagnetic modes of the cavity or package itself give rise to several peaks in ${\rm{Re}}~Z_{\rm{env}}\left[\omega\right]$, that can be extracted from an electromagnetic simulation of the entire package~\cite{nigg_bbq_2012, minev2021_EPR}.
We show the response associated with the readout mode (EM0) and the other spurious electromagnetic modes (EM1-EM9) in Fig.~\ref{fig:motivation}(f) of our package as an example of such an analysis. 

To reveal the DUST mechanisms in our experimental device, we perform a time-resolved pump-probe spectroscopy on the fixed-frequency transmon. In this experiment, we initialize the transmon in either the ground state or the first excited state and apply a pump pulse of variable frequency and power to it. We then measure the final state populations of the transmon to quantify the probability of unwanted transitions. 
The resulting transition probabilities exhibit several resonant features that occur under specific drive conditions. 
We categorize these observed features into the three DUST mechanisms (\textit{A}-\textit{C}) through a sequence of checks: 
Features that do not show an explicit dependence on the drive frequency but solely depend on the induced ac-Stark shift are attributed to mechanism \textit{A}, resonant exchange with a spurious mode. 
In contrast, both mechanisms \textit{B} and \textit{C} result in features that have both drive frequency and power dependence. 
To further discriminate between mechanism \textit{B} and \textit{C}, we perform a driven steady-state simulation of a model Hamiltonian containing only the transmon mode. 
By comparing the simulation results with the experimental transition map, we identify all the features arising from mechanism \textit{B}, intrinsic multi-photon excitations. 
Finally, we attribute the remaining features to mechanism \textit{C}, inelastic scattering involving external modes. 
By investigating the frequency dependence of these features, we pinpoint the environment modes involved in such inelastic scattering processes. The majority of the features are explained by spurious RF modes predicted by a finite-element electromagnetic simulation of the experimental device. 
We attribute the remaining few features to inelastic processes involving material defects with unpredictable frequencies.

Our experimental data show that, for a given device, certain drive frequencies are less susceptible to DUST. Such frequencies can be chosen for engineering readout or control drives to maximize the fidelity of these operations. 
We underscore that most of the observed transitions are predictable through a driven transmon simulation and electromagnetic environment simulation. 
These simulations should be a guide for device design and frequency allocation in experiments with superconducting circuits. 
The mechanisms drive-induced unwanted state transitions that we uncover, as well as the diagnostic spectroscopy technique that we develop, are relevant to all superconducting non-linear circuits, including fluxonium~\cite{manucharyan2009, floquet_fluxonium}, multimodal circuits~\cite{remy_2020, hazra_2025, chapple2024, chapple2025}, SNAIL~\cite{frattini_2017}, and various couplers~\cite{sete_2021, lu_2023, ats_2023,inductive_coupler_2024, maiti2025}.

This article is organized as follows. Section~\ref{sec:dust_spectroscopy} describes the results of the time-resolved pump-probe spectroscopy experiment. Sections~\ref{sec:tls_ac_stark_shift} and \ref{sec:intrinsic_mpt} investigate mechanisms \textit{A} and \textit{B}, respectively. Section~\ref{sec:floquet_hybridization} demonstrates how all the intrinsic multi-photon transitions (mechanism \textit{B}) are identified through Floquet simulations. Section~\ref{sec:geometric_inelastic} illustrates the experimental evidence for mechanism \textit{C}. Section~\ref{sec:extrinsic_multi_photon} presents a two-mode Floquet analysis simulating a higher-order inelastic scattering. 
Section~\ref{sec:inelastic_tls} highlights an experimentally observed inelastic process assisted by a TLS. Finally, Section~\ref{sec:conclusions} concludes the article by summarizing the results and discussing the design strategies to suppress such unwanted transitions.

%%=============================================================
%% SPECTROSCOPY OF DRIVE-INDUCED UNWANTED STATE TRANSITIONS
%%=============================================================
%%-------------------------------------------------------------
%% Figure 2: DUST in a fixed frequency 3D transmon.
%%-------------------------------------------------------------
\begin{figure*}[t!]
\includegraphics[width = 0.96\textwidth]{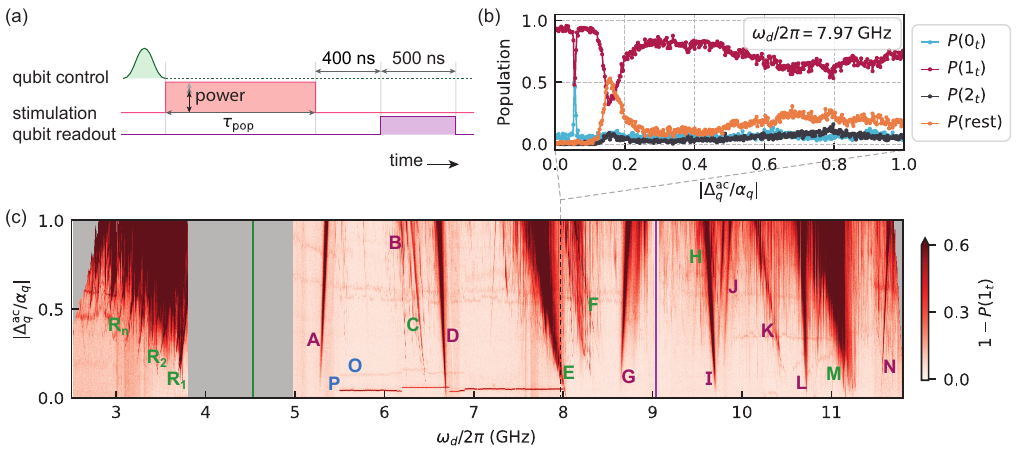}
\caption{DUST in a fixed frequency 3D transmon. (a) Pulse sequence for the experiment. The transmon is prepared in the excited state with a $\pi$ pulse followed by a $\tau_{\rm{pop}} = 1~\mu$s stimulation pulse. The frequency $\omega_d$ and the power $\bar{n}_r$ of the pulse are swept. The final state of the transmon is measured by single shot readout after each experiment.
(b) Example of unwanted transitions from $\ket{1_t}$ of the transmon when it is driven at a frequency $\omega_d/2\pi = 7.97$ GHz. Final populations of transmon state are plotted as a function of the drive power. The drive power is calibrated through the drive induced ac-Stark shift on the qubit, normalized by the transmon anharmonicity.  
(c) Transitions from  $\ket{1_t}$ of the transmon measured by the  two-tone spectroscopy with variable drive power and frequency.
The color plots show the transition probability from the $\ket{1_t}$ of the transmon at the end of the stimulation pulse. 
Data for $\ket{0_t}$ initialization is shown in Appendix ~\ref{sec:landscape_g_and_e}.
The plot reveals multiple resonant features, showing distinct transitions at particular combinations of drive frequency and power. We label the prominent transitions above the qubit frequency with Roman letters \textbf{A} through \textbf{P}. For drive frequencies below the qubit transition, the resonances appear as densely packed features. We do not assign individual Roman letters to these transitions and instead label them as \textbf{R\textsubscript{1}}, \textbf{R\textsubscript{2}}, …, \textbf{R\textsubscript{n}}. The blue labels (\textbf{O} and \textbf{P}) correspond to resonant exchange with spurious modes. The green and magenta labels correspond to intrinsic multi-photon resonances and inelastic scattering processes, respectively.
The qubit and the readout frequency are shown by the vertical green and violet lines, respectively. 
}
\label{fig:landscape}
\end{figure*}
%%-------------------------------------------------------------

\section{Spectroscopy of drive-induced unwanted state transitions}
\label{sec:dust_spectroscopy}
We perform the time-resolved pump-probe spectroscopy on a fixed-frequency transmon with frequency $\omega_{q}/2\pi = 4.5285$ GHz and anharmonicity $\alpha_q/2\pi = -184.2$ MHz. 
The pump is sent through a dedicated drive port. The filtering and cryogenic attenuation of the input drive line are engineered for power delivery across a wide frequency span, ranging approximately from $2.8$ GHz to $11.8$ GHz, while maintaining the qubit coherence. 
The readout resonator has a frequency $\omega_r/2\pi = 9.0342$ GHz and a linewidth $\kappa/2\pi = 7.20$ MHz. 
The qubit state dispersive shift of the resonator is $\chi/2\pi = -1.23$ MHz. The dispersive shift is chosen to be smaller than the readout resonator linewidth to resolve multiple transmon states through a single-shot readout. The details of the device parameters and the readout performance are discussed in Appendix~\ref{sec:device}.
In the pump-probe experiment, we prepare the transmon in either $\ket{0_t}$ or $\ket{1_t}$ and apply a $1~\mu$s long stimulation drive of varying frequency and power. 
Then we add a wait time of $400$ ns for any residual photons to leave the cavity. 
Finally, we perform a readout of the transmon to measure the final populations of the transmon states, $P(0_t)$, $P(1_t)$, $P(2_t)$, and $P({\rm{rest}})$, where $P({\rm{rest}}) = 1-P(0_t)-P(1_t)-P(2_t)$. The pulse sequence of the experiment is shown in Fig.~\ref{fig:landscape}(a).

The displacement of the transmon induced by the stimulation drive~\cite{Xiao2023} depends on the detuning of the drive from the qubit frequency.
Moreover, the effective drive power reaching the transmon is filtered by a frequency-dependent transfer function. 
To ensure a consistent calibration across the full drive-frequency range, we quantify the drive power in terms of the induced ac-Stark shift $\Delta_q^{ac}$, on the transmon. 
At each frequency, we fit the measured $\Delta_q^{ac}$ versus applied power to extract a calibration slope, which we then use to rescale the entire power sweep from $\Delta_q^{ac} = 0$ to $\Delta_q^{ac} = \alpha_q$, where $\alpha_q$ is the transmon anharmonicity.

As an example, we show the result of this experiment for a given drive frequency, $\omega_d/2\pi = 7.97$ GHz, in Fig.~\ref{fig:landscape}(b).  In this experiment, we prepare the transmon in $\ket{1}$, and plot the final transmon populations as a function of drive power.  At zero drive strength, the $\ket{1_t}$ population is $93\%$, limited by state preparation and measurement (SPAM) errors and natural decay during the wait time.  When the induced ac-Stark shift equals $|\Delta_q^{\rm{ac}}/\alpha_q| = 0.055 $, the transmon hits a resonance and it rapidly decays to $\ket{0_t}$. At $|\Delta_q^{\rm{ac}}/\alpha_q| = 0.155 $, a second resonance excites the transmon to a non-computational state.
At even higher powers, we observe a broadened feature, primarily exciting the transmon to non-computational states, along with relatively smaller population transfers to $\ket{0_t}$ and $\ket{2_t}$ at specific drive powers.

In Fig.~\ref{fig:landscape}(c), we plot the total transition probability, $1-P(1_t)$, as a function of both the pump power and frequency in the pump-probe spectroscopy experiment. The gray regions show the parameter space where we interrupted the sweep as the power exceeds the linear regime of the room temperature setup (above 11.5 GHz and below 3 GHz). For visual clarity, we also interrupt the sweep where the drive frequency approaches the qubit frequency (marked by the green line), which would otherwise activate off-resonant Rabi oscillations either between $\ket{0_t}$ and $\ket{1_t}$ or $\ket{1_t}$ and $\ket{2_t}$ \footnote{In the part of the excluded region the sign of the Stark shift changes in the straddling regime of the transmon~\cite{transmon}, further complicating the power calibration.}. The experimental results reveal a landscape of state transitions as a function of drive frequency $\omega_d$ and drive power.  These transitions exhibit distinct resonant features, with the transition probabilities peaking at specific combinations of frequency and power. To refer to specific transitions in this plot, we label them with roman letters.

Most resonances appear as lines with positive or negative slopes in the $(\omega_d, \Delta_q^{ac}$) plane, representing decay and excitation processes, respectively. In addition, we observe \textit{quasi-horizontal} resonant features that depend only on the induced ac-Stark shift, and not explicitly on the drive frequency. We label two such transitions by \textbf{O} and \textbf{P} in Fig.~\ref{fig:landscape}(c). A dense set of transitions, labeled \textbf{R\textsubscript{1}}, \textbf{R\textsubscript{2}}, \ldots, \textbf{R\textsubscript{n}}, appears when the drive frequency is below $\omega_q$. With increasing drive power, the sloped resonances broaden in frequency. Some transitions, such as \textbf{E}, \textbf{F}, and \textbf{M} broaden more significantly than others (e.g., \textbf{A}, \textbf{D}, \textbf{G}, and \textbf{I}). In addition,  these broadened transitions also exhibit temporal fluctuations, giving them a ``fuzzy'' appearance at large drive powers. 

We observe similar resonant features when we prepare the transmon in the ground state (see Appendix~\ref{sec:landscape_g_and_e} for details).
For the remainder of the article, we focus on the underlying mechanisms responsible for these transitions and classify all labeled transitions into distinct categories. For each category, we discuss design strategies that can predict and suppress these transitions.

%%=============================================================
%% AC-STARK-SHIFT-INDUCED RESONANT EXCHANGE WITH A SPURIOUS MODE
%%=============================================================

\section{ac-Stark-shift-induced resonant exchange with a spurious mode}
\label{sec:tls_ac_stark_shift}
%%-------------------------------------------------------------
%%% Figure 3: Qubit brought into resonance with a TLS 
%%-------------------------------------------------------------
\begin{figure}[t]
\includegraphics[width = 0.48\textwidth]{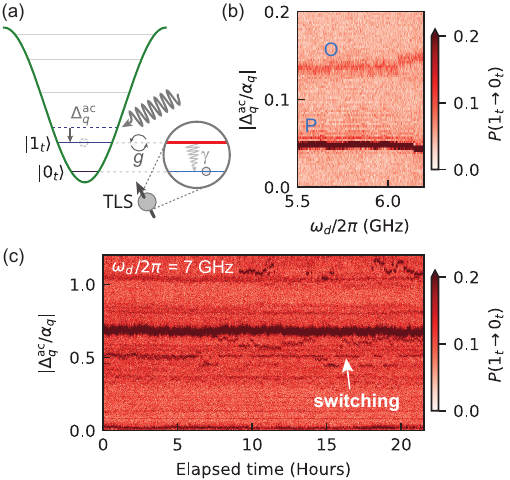}
\caption{Qubit brought into resonance with a TLS by ac-Stark shift.  
(a) The transmon experiences a drive-induced ac-Stark shift, bringing the shifted $\ket{1_t}$ level into resonance with a dissipative TLS, relaxing the transmon into the system’s thermal ground state. (b) Magnified section of Fig.~\ref{fig:landscape}(c) to highlight two prominent features \textbf{P} and \textbf{O} showing a transmon relaxation due resonant exchange with TLS. (c) Temporal change of the TLS environment, probed through the decay probability of the transmon, as a function of normalized ac-Stark shift. The TLS bath exhibits both quasi-continuous drift and telegraphic-noise-like switching when monitored over $\sim 22$ hours. 
The frequency of the drive inducing the ac-Stark shift is chosen such that the drive does not activate other detrimental transitions.
}
\label{fig:tls_spectrum} 
\end{figure}
%%-------------------------------------------------------------

We first focus on the \textit{quasi-horizontal} features in Fig.~\ref{fig:landscape}(c). 
These features arise from interactions of the transmon with a dissipative bath of discrete modes, such as two-level systems (TLS). In this article, we refer to any saturable, spurious degree of freedom of unknown origin that exhibits slow temporal drift or discrete switching as a TLS\cite{Muller2019_TLS}.
The static transmon-TLS interaction is usually modeled as a transverse  bilinear coupling.
When the dressed transmon frequency, $\tilde{\omega}_{q} = \omega_{q}+\Delta_q^{\rm{ac}}$, becomes resonant with the frequency of a TLS, the transmon exchanges energy with the TLS at a rate given by their transverse interaction strength~\cite{Cole2010_TLS}. 
If the TLS is cold and dissipative, this leads to decay of the transmon from $\ket{1_t}$ to $\ket{0_t}$. 
Such processes are known to limit qubit $T_1$ times~\cite{Klimov2018, burnett2019_TLS, Sivak2023_gkp, Thorbeck_2024_readout_T1}. If the TLS is thermally populated, it can also induce excitations in the transmon upon a resonant exchange. Furthermore, when the coupling strength $g$ is comparable to or exceeds the TLS dissipation rate $\gamma$, coherent energy swapping can occur, leading to non-Markovian dynamics between the transmon and its environment~\cite{spiecker2023, Chen2024_TLS}.

In Fig.~\ref{fig:tls_spectrum}(b) we zoom into the transitions labeled \textbf{P} and \textbf{O} in the previous Fig.~\ref{fig:landscape}(c). 
Transition \textbf{P} displays coherent dynamics when the transmon is tuned into resonance with a TLS. We analyze this interaction in more detail in Appendix\ref{sec:tls_decay_dynamics}.
The apparent slow drift of the transition as we change the drive frequency is caused by the temporal drift of the TLS frequency itself. 

To further probe the temporal frequency drift of the TLS environment, we utilize the ac-Stark shift as a knob to sweep the transmon frequency. 
We fix the drive frequency to be $\omega_d/2\pi = 7$ GHz where we can tune the qubit frequency by more than $200$ MHz, without causing any unwanted transitions arising from the other two mechanisms. This \textit{silent} window is chosen by inspecting the landscape in Fig.~\ref{fig:landscape}(c). 
This technique is analogous to flux tuning a SQUID transmon~\cite{Klimov2018,Lisenfeld2019_TLS,Chen2024_TLS}, or a fluxonium~\cite{gunzler2025} in order to characterize its environment. 

In this experiment, we initialize the transmon in $\ket{1_t}$ and turn on the drive for $5~\mu$s followed by a readout to measure the population transfer $P(1_t\rightarrow0_t)$. 
Note that we have increased the drive duration from $1~\mu$s in the previous experiment to increase the contrast and identify weakly coupled TLS in the environment. 
We observe several distinct loss peaks (\textit{hot-spots}) at particular values of the ac-Stark shift, at which the transmon tunes into resonance with an individual TLS. 
We monitor these TLS hot-spots for about $22$ hours to investigate their temporal fluctuations as shown in Fig.~\ref{fig:tls_spectrum}(c). 
The temporal change in the TLS environment consists of several different timescales ranging from minutes to hours. Some TLS exhibit a quasi-continuous frequency drift, while others show a telegraph-noise-like switching between well-defined frequencies. In addition, some TLS appear and disappear from the spectroscopic range in the span of several hours. 
These observations are consistent with previously reported behavior of TLS ~\cite{Martinis2005_TLS, Muller2015_TLS, Klimov2018, burnett2019_TLS}.

%%=============================================================
%% INTRINSIC MULTI-PHOTON RESONANCES
%%=============================================================
\section{Intrinsic multi-photon resonances}
\label{sec:intrinsic_mpt}

Most features in Fig.~\ref{fig:landscape}(c) appear as sloped lines, some of which exhibit additional curvature at higher powers. 
A subset of these transitions is explained by mechanisms intrinsic to the ideal transmon circuit itself, i.e. those captured by the driven transmon Hamiltonian,
\begin{equation}
\label{eq:Hamiltonian}
H = 4E_{C}(\hat{n}-n_{g})^2 - E_{J}\cos(\hat{\varphi}) + E_d \hat{n} \cos(\omega_d t),
\end{equation}
where $E_C$ is the charging energy of the transmon, $E_J$ is the Josephson energy, $\hat\varphi$ and $\hat{n}$ are respectively the phase and charge operators, and $n_g$ is the offset charge. We denote the eigenstates and eigenenergies of Eq.~\eqref{eq:Hamiltonian} as $\ket{i_{t}}$, $\omega_{i}$, respectively, and define the transition frequency $\omega_{ji}\equiv\omega_{i}-\omega_{j}$.

\begin{figure}[t]
\includegraphics[width = 0.48\textwidth]{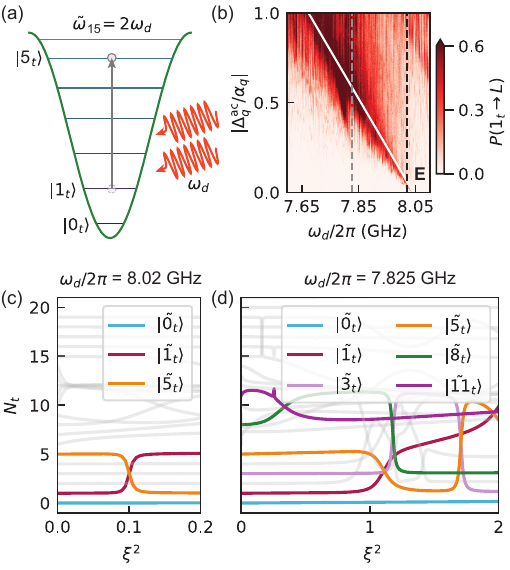}
\caption{Intrinsic multi-photon transition to non-computational levels of the transmon. (a) A schematic showing a resonance condition that annihilates two drive photons to excite the transmon from $\ket{1_t}$ to $\ket{5_t}$ 
(b) A zoom-in of the measured $P(1_t\rightarrow L$) transition, highlighting the feature \textbf{E}.
The white solid line represents the driven resonance condition, $\tilde{\omega}_{15} \approx {\omega}_{15} +4\Delta_q^{\rm{ac}}=2\omega_d$, from the leading order perturbation theory. Although the theory shows decent agreement at low power, it deviates at a higher drive power. 
(c) Branch analysis~\cite{Dumas_2024_Ionization} of the transition, showing a branch swap between $\ket{\tilde{1}_t}$ and $\ket{\tilde{5}_t}$ for $\omega_{d}/2\pi=8.05$ GHz at a relatively low power, $|\xi|^2 = 0.1$. (d) When driving at $\omega_{d}/2\pi=7.825$ GHz, the same transition as in (c) now occurs at a higher drive power, $|\xi|^2 \approx 1.1$, due to the ac-Stark shift. An additional branch-swapping event occurs at $|\xi|^2 = 1.7$, involving the state $\ket{\tilde{11}_t}$ that has itself experienced multiple prior branch swaps. 
At high drive powers, several such transitions become activated and hybridize with each other. The corresponding resonance conditions are sensitive to offset charge, 
resulting in a broad, fuzzy feature in the ensemble-averaged experimental data.
}
\label{fig:intrinsic_resonance}
\end{figure}

To illustrate an intrinsic multi-photon excitation, we focus on the transition labeled \textbf{E} in Fig.~\ref{fig:landscape}(c) . This feature corresponds to a resonant process that annihilates two drive photons to excite the transmon from state $\ket{1_t}$ to state $\ket{5_t}$ in a \textit{six-wave mixing} process, as shown in Fig.~\ref{fig:intrinsic_resonance}(a). 
We can follow a perturbative approach to understand the resonance condition for this process~\textemdash both where it occurs in drive-frequency space, and why it appears as a sloped line in the drive-frequency vs. drive-power plot. 
The undriven energy difference between $\ket{1_t}$ and $\ket{5_t}$ is experimentally found to be $\omega_{15}/2\pi = 16.084$ GHz. 
Thus, in the weak drive limit, the resonance condition is met when the drive is applied at $\omega_d/2\pi=(\omega_{15}/2\pi)/2 = 8.042$ GHz. 
As the drive power increases, the transmon levels experience ac-Stark shift; within the quartic approximation, this is given by $\Delta_{\ell}^{\rm{ac}}=\ell \Delta_{q}^{\rm{ac}}$ for level $\ket{\ell_{t}}$, where $\Delta_{q}^{\rm{ac}}=\xi^2 \alpha_q/2$
is the ac-Stark shift of the qubit-transition frequency, and
\begin{equation}
\label{eq:xi}
\xi = \frac{2 n_{\ZPF} \omega_d}{\omega_d^2 - \omega_{01}^2} \frac{E_d}{\hbar},
\end{equation}
is a dimensionless number measuring the displacement of the transmon mode, and $n_{\ZPF}$ is the zero-point fluctuation of the charge operator. See Appendix~\ref{sec:xi} for a derivation of the expressions for $\xi$ and $\Delta_{q}^{\rm{ac}}$.

The transition frequency therefore approximately shifts to  $\tilde{\omega}_{15} = \omega_{15}+4\Delta_q^{ac}$ in the presence of the drive. Thus, to leading order, the resonance condition is $\omega_d = \omega_{15}/2+2\Delta_q^{ac}$, shown as a white line in Fig.~\ref{fig:intrinsic_resonance}(b). This simple expression is in good agreement with the experimental data at low drive powers. As the drive power increases, higher-order interactions and cascaded processes also become relevant, and the main transition further hybridizes with several secondary transitions. In this case, a higher-order perturbation theory becomes necessary, including all relevant processes as well as non-RWA effects as shown in Ref.~\cite{Xiao2023}. However, previously developed perturbative approaches involve an explicit expansion of the cosine potential, and thus cannot capture the effects of offset-charge dependence of these transitions.

Floquet steady-state simulations provide a natural alternative for analyzing such strongly driven systems. These simulations account for the periodic nature of the drive and can accurately predict the intrinsic multi-photon resonances in the driven transmon Hamiltonian, given by Eq.~\ref{eq:Hamiltonian}. Unlike perturbation theory, the Floquet formalism does not require expansion of the cosine potential or truncation of the order of interaction, and has been previously applied to explain and predict the behavior of driven superconducting circuits~\cite{Wang2025, Dumas_2024_Ionization, Fechant2025, Kurilovich2025}. 
These resonances can be visualized by performing a so-called {\it branch analysis} \cite{Dumas_2024_Ionization, Shillito2022, Cohen2023}, where the Floquet modes are tracked as a function of the drive strength, see Appendix~\ref{app:branch_analysis} for details. 
In the absence of multi-photon resonances, the resulting branches appear as nearly horizontal lines (with small slopes indicating ac-Stark shift). When a multi-photon resonance occurs, two branches swap.

%%-------------------------------------------------------------
%%% Figure 4: Intrinsic multi-photon transition
%%-------------------------------------------------------------
%%-------------------------------------------------------------

In Fig.~\ref{fig:intrinsic_resonance}(c), we show the branch analysis performed at drive frequency $\omega_d/2\pi = 8.02$ GHz, marked with the dashed black line in Fig.~\ref{fig:intrinsic_resonance}(b). The branch swap between $\ket{\tilde1_t}$ and $\ket{\tilde5_t}$ further confirms the intrinsic multi-photon transition process. 
We perform another branch analysis at a smaller drive frequency $\omega_d/2\pi = 7.825$ GHz, at which the resonance condition is met at a higher drive power (see the gray dashed line in Fig.~\ref{fig:intrinsic_resonance}(b)). We find that at stronger drive powers, the state $\ket{\tilde5_{t}}$ further hybridizes with another non-computational state, $\ket{\tilde{11}_t}$, through several branch swaps.

In the Floquet simulation, the branch swaps occur at particular drive powers. Thus, the resonances are expected to appear like sharp lines, up to the power broadening of these transitions. In experiments, however, we observe that the broadening of these transitions is not solely explained by power broadening. 
We understand the smearing and the ``fuzziness'' of the transition at higher power once we take into account the drifting offset charge ($n_g$) of the transmon island. 
As $n_g$ drifts over the timescale of a single experiment, the resonance condition shifts accordingly~\cite{Sank_2016_MIST, Khezri_2023_MIST, Fechant2025}. 
Consequently, the measured population, averaged over many single-shot readouts, appears fuzzy or smeared due to this temporal offset charge variation. In addition, when the drive activates resonances between one non-computational transmon state to another, it connects $\ket{1_t}$ to multiple final states, each with its own $n_g$ dependence. 
The charge dispersion of these levels scales exponentially with the transmon excitation number \cite{transmon, schreier_transmon_2008}.
Thus, we expect a stronger dispersion of the resonance condition at stronger drive powers as multiple non-computational states participate in the transition.
Consequently, the observed signal reflects a mixture of these overlapping, charge-sensitive states. See Appendix~\ref{app:ng_drift} for a detailed discussion on the $n_g$ sensitivity of these transitions.

%%=============================================================
%% IDENTIFYING INTRINSIC TRANSITIONS AT DIFFERENT FREQUENCIES
%%=============================================================

\section{Identifying intrinsic transitions at different frequencies}
\label{sec:floquet_hybridization}
%%-------------------------------------------------------------
%%% Figure 5: Identification of intrinsic multi-photon transitions
%%-------------------------------------------------------------
\begin{figure}[t]
\includegraphics[width = 0.48\textwidth]{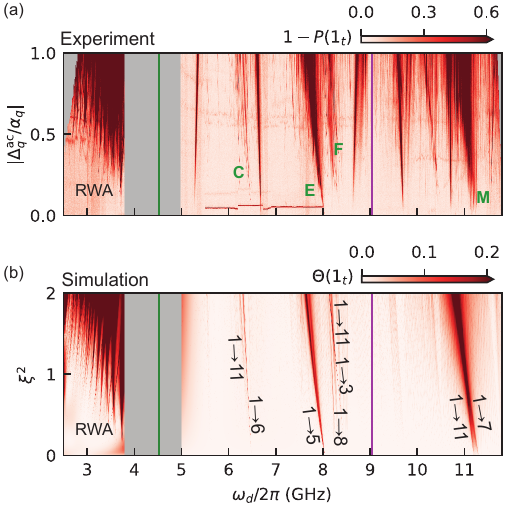}
\caption{Identification of intrinsic multi-photon transitions through Floquet simulations. (a) Experimental data showing transitions from $\ket{1_t}$, same as Fig.~\ref{fig:landscape}(c), with the labels highlighting only the intrinsic multi-photon transitions. (b) Floquet steady-state simulation result showing the hybridization $\Theta(1_{t})$ between the driven computational state $|\tilde{1}_t\rangle$ with driven non-computational states. Plotted results are averaged over a uniform distribution of $n_{g}$ values.
}
\label{fig:floquet_theory} 
\end{figure}
%%-------------------------------------------------------------

We identify all intrinsic multi-photon transitions in Fig.~\ref{fig:landscape}(c) by performing a Floquet simulation for the entire frequency range of the pump-probe spectroscopy experiment. While the branch-analysis technique provides a systematic approach for labeling transitions, it becomes visually complicated when the drive frequency and offset charge are swept. Therefore, to compare the experimental data with Floquet simulation, we compute the \textit{hybridization parameter}~\cite{Xiao2023, Kurilovich2025}
\begin{align}
\Theta(j_t) = 1 - |\langle \tilde{j}_{t}(\xi,\omega_d)|\bar{j}_t(\xi,\omega_d)\rangle|^2,
\end{align}
which quantifies the degree of hybridization with higher-lying states. Importantly, $\Theta(j_{t})$ calibrates out any hybridization due only to drive-induced ac-Stark shift through the definition of the {\it ideal-displaced state} $|\bar{j}_t(\xi,\omega_d)\rangle$~\cite{Xiao2023} (see Appendix~\ref{sec:floquet_sim} for details on the definition of $|\bar{j}_t(\xi,\omega_d)\rangle$, and the procedure for assigning transmon labels to the Floquet modes).
Thus,  $\Theta(j_{t})$ remains close to zero in the absence of multi-photon resonances, when the Floquet states are well approximated by the displaced transmon eigenstates. 
However, $\Theta(j_{t})$ experiences a sharp increase when the drive condition satisfies a multi-photon resonance involving a non-computational state. 

To detect all intrinsic transitions that arise from the transmon Hamiltonian itself, we compute $\Theta(1_{t})$ over the same range of frequencies as in Fig.~\ref{fig:landscape}(c), see Fig.~\ref{fig:floquet_theory}(b) [results for $\Theta(0_{t})$ are shown in Appendix~\ref{sec:landscape_g_and_e}].
The range of the drive amplitudes is chosen such that the induced ac-Stark shift is consistent with the experiment. 
We conduct the Floquet simulation with randomized $n_g$, and the averaged $\Theta(1_t)$ is plotted. The transition lines broaden as the drive power increases, due both to the strength of the processes scaling with drive power, as well as the $n_{g}$ sensitivity of the resonance conditions. By averaging over $n_{g}$, we mimic the experiment where the offset charge is an unknown parameter that is typically quasi-static on the timescale of a single shot, but fluctuates between shots~\cite{Fechant2025}. 
The Floquet simulations predict the experimentally observed drive conditions 
 for activating multi-photon transitions for multiple features shown in Fig.~\ref{fig:floquet_theory}(a).

We label the transitions above the qubit frequency $\omega_q$ in Fig.~\ref{fig:floquet_theory}(b) by performing a branch analysis with $n_{g}=0.25$. As in Fig.~\ref{fig:intrinsic_resonance}(c), this is generally done at a frequency only slightly negatively detuned from the bare-resonance condition, for ease in identifying the state(s) involved.
As introduced in Sec.~\ref{sec:intrinsic_mpt}, the branch analysis reveals that multiple levels can be hybridized simultaneously at certain frequencies, leading to complicated transitions beyond the simple crossing of two levels. For instance, the rightmost feature in Fig.~\ref{fig:floquet_theory}(b) is a resonance between transmon states $\ket{\tilde1_{t}},\ket{\tilde7_{t}}$ and $\ket{\tilde11_{t}}$.
For drive frequencies below $\omega_q$, we observe a dense set of transitions into leakage states. These features arise due to the transmon’s weakly negative anharmonicity, which results in a ladder of closely spaced transitions from $\ket{1_t}$ to higher levels, $\ket{(1+n)_t}$ absorbing $n$ drive photons. This set of excitation-number preserving multi-photon resonances has been previously reported~\cite{Khezri_2023_MIST} in a driven transmon.
Our data demonstrates the predictive power of the  Floquet simulation in identifying the drive conditions for intrinsic multi-photon resonances. 
In principle, from the Floquet simulation, one can also extract the rate of these transitions for a given drive condition.
However, in this work, we have not explicitly connected the hybridization parameter $\Theta(j_t)$ and the experimentally measured transition probability. We provide the code used for these simulations as the open-source package \texttt{floquet} \cite{floquet2024}.

Note that the accuracy of the Floquet simulation predictions depends crucially on the underlying Hamiltonian. First, any deviation of the transmon potential from the cosine shape caused by either higher Josephson harmonics in tunnel junctions or stray lead inductance will modify the spectrum~\cite{willsch_harmonics} and thus the onset of these transitions. Second, any spurious degrees of freedom that participate in the junction will activate additional transitions. 
Indeed, we observe that only a handful of transitions recorded in the experimental spectrum are reproduced in the Floquet simulation based on the Hamiltonian in Eq.~\ref{eq:Hamiltonian}.
In the following sections, we elucidate the physical origins of these ``extrinsic'' DUST mechanisms and develop a phenomenological framework to predict and mitigate their impact in driven quantum circuits.

%%=============================================================
%% INELASTIC SCATTERING INVOLVING SPURIOUS MODES
%%=============================================================
\section{Inelastic scattering involving spurious modes}
\label{sec:geometric_inelastic}

In the previous section, the static circuit Hamiltonian for the Floquet simulation includes only the isolated transmon consisting of a Josephson junction and a shunting capacitor. However, in a realistic system, the transmon is unavoidably coupled to a more complex environment. This includes the fundamental mode and higher harmonics of the readout circuitry~\cite{houck_2008}, as well as spurious package modes. Such a coupling leads to additional DUST mechanisms.
Indeed, the experimental data reveal several resonance-like transitions arising from inelastic scattering processes involving the spurious electromagnetic modes. 

%%-------------------------------------------------------------
%%% Figure 6: Inelastic scattering processes (EM mode)
%%-------------------------------------------------------------
\begin{figure}[t]
\includegraphics[width = 0.48\textwidth]{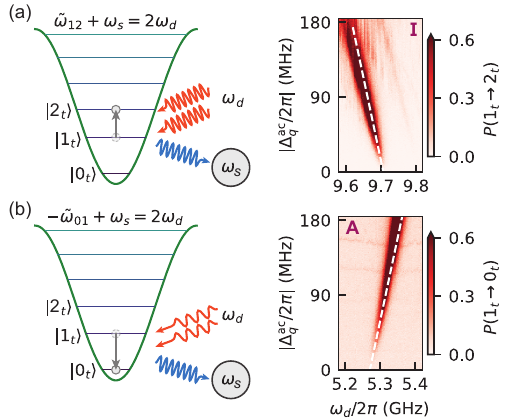}
\caption{Inelastic scattering processes assisted by a spurious electromagnetic mode.
(a) (left) A two-mode squeezing process activated by the drive, exciting the transmon from $\ket{1_t}$ to $\ket{2_t}$, and simultaneously exciting the spurious mode at $\omega_s/2\pi = 15.07$ GHz. (right) Corresponding transition, identified in Fig.~\ref{fig:landscape}(c), labeled by \textbf{I}. The white dashed line shows the expected resonance condition $\tilde{\omega}_{12} \approx {\omega}_{12}+\Delta_q^{\rm{ac}} = 2\omega_d - \omega_s$ from a leading order perturbation theory.
(b) (left) A two-mode conversion process, mediated by the same spurious mode, activated at a different drive frequency. This process absorbs two drive photons, annihilates one quantum of energy from the transmon causing it to decay from $\ket{1_t}$ to $\ket{0_t}$, and releases the energy into the spurious mode at $\omega_s$. (right) This transition is identified with the label \textbf{A} in Fig.~\ref{fig:landscape}(c) and the white dashed line represents the resonance condition $-\tilde{\omega}_{01} = -{\omega}_{01} -\Delta_q^{\rm{ac}}=2\omega_d - \omega_s$.
}
\label{fig4:heating_decay} 
\end{figure}
%%-------------------------------------------------------------

Such an inelastic scattering process either annihilates $n$ photons from the drive and excites the transmon $\ket{i_t} \rightarrow \ket{j_{t}}$, or causes the transmon to relax $\ket{1_t} \rightarrow \ket{0_t}$. The excess energy is emitted into the environment in the form of $n$ photons at a frequency $\omega$.
The strength of this process depends on 
the transmon environment $\rm{Re}\left(Z\right[\omega])$, and is enhanced whenever there is a resonant mode coupled to the transmon at the emission frequency $\omega_s$~\cite{connolly_2025}.
As an example, in Fig.~\ref{fig4:heating_decay}, we show two such inelastic scattering processes, mediated by a spurious mode at $\omega_s/2\pi = 15.07$ GHz. This mode is independently confirmed in experiment (see Appendix \ref{sec:extrinsic_hfss} for details), and is further identified as the $\rm{TE}103$ mode of the readout cavity from electromagnetic simulations. Fig.~\ref{fig4:heating_decay}(a) shows a \textit{two-mode squeezing process} that annihilates two drive photons to excite the transmon from $\ket{1_t}\rightarrow\ket{2_t}$ and releases the excess energy into the dissipative spurious mode. 
As the drive power increases, the transmon levels experience ac-Stark shift, thus meeting the resonance condition  at a lower drive frequency. The resonant feature therefore has a negative slope in the spectroscopy plot, similarly to the intrinsic processes described in Sec.~\ref{sec:intrinsic_mpt}.
This mode can also mediate a \textit{two-mode conversion process} that absorbs two drive photons, annihilates a qubit excitation, and emits a photon into the dissipative spurious mode [see Fig.~\ref{fig4:heating_decay}(b)]. 
As the qubit frequency decreases with increasing drive power due to the ac-Stark shift, the resonance condition is satisfied at higher drive frequencies. Consequently, the resonant feature in the spectroscopy plot has a positive slope. The white dashed line in each plot represents the expected resonance condition from leading order perturbation theory, and is in agreement with the experimental data at lower drive powers. 

In the section above, we have discussed only single quantum excitations and relaxations mediated by spurious electromagnetic modes. However, this mechanism additionally 
introduces channels for multi-quanta excitations of the transmon through higher-order mixing processes. 
We discuss these processes in detail in Sec.~\ref{sec:extrinsic_multi_photon}. In general, for each spurious mode coupled to the transmon, a transition can occur if the resonance condition $\left(\tilde\omega_{i(i+\ell)}+m \omega_s = n\omega_d,\;\forall\{\ell,m,n\}\in \mathbb{Z}\right)$
is satisfied, where $\omega_s$ is the frequency of the spurious mode. The symmetry of the transmon potential ideally imposes a selection rule for these transitions, allowing only those where $(\ell+m+n)$ is even. 
%~\footnote{Non-zero offset charge breaks this selection rule. But these symmetry-breaking transitions are still suppressed if the resonance only involves first few transmon states}
Note that this selection rule can be broken under non-zero offset charge. \cite{Fechant2025} To identify all the modes that participate in such processes, we perform a finite element electromagnetic simulation of the package up to $30$ GHz. 
These simulated eigenmodes explain the family of transitions \textbf{A}, \textbf{D}, \textbf{G}, \textbf{I}, \textbf{J}, and \textbf{N}, within a leading-
order perturbation theory and a quartic approximation of the transmon potential.
We label these transitions by bold magenta letters in Fig.~\ref{fig:landscape}(c). See Appendix~\ref{sec:extrinsic_hfss} for details on how we identify and label these transitions through RF simulations.
Our findings underscore that package modes, even far above the frequency of the qubit, can contribute significantly to DUST.  For example, the package mode EM9 around $28$ GHz is responsible for the resonant feature \textbf{N} in Fig.~\ref{fig:landscape}(c). Hence, it is important to engineer the frequencies of these parasitic modes and simulate their impact on the driven circuit while designing the layout.

%%=============================================================
%% HIGHER ORDER INELASTIC SCATTERING PROCESSES
%%=============================================================
\section{Higher order inelastic scattering processes}
\label{sec:extrinsic_multi_photon}

%%-------------------------------------------------------------
%%% Fig 7: Inelastic scattering process, multi-excitation
%%-------------------------------------------------------------
\begin{figure}
    \centering
    \includegraphics[width=0.48\textwidth]{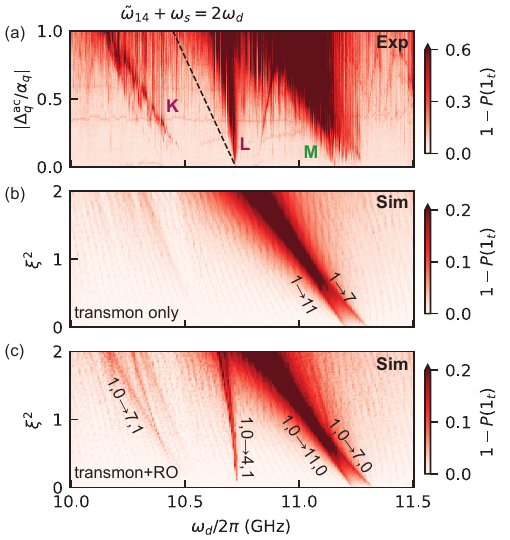}
    \caption{Inelastic scattering process causing multi-excitation of the transmon. (a) Experimental data and (b) its comparison with a transmon-only Floquet simulation. The features labeled \textbf{K} and \textbf{L} do not appear in the Floquet simulation, suggesting that they involve excitations of an external mode. The dashed black line in (a) is an expected resonance condition involving the transmon and the readout resonator, assuming a quartic approximation of the transmon potential and which strongly deviates from the data at higher power. (c) Floquet simulation of the driven Hamiltonian that includes the transmon and the readout resonator (RO). In addition to the intrinsic transitions within the transmon circuit, the simulation accurately predicts the joint excitations of the transmon-resonator system. 
    }
    \label{fig:two_mode_floquet}
\end{figure}
%%-------------------------------------------------------------

Although a leading-order perturbation theory predicts most of the observed inelastic scattering processes that involve a spurious mode, some transitions remain unexplained.  
This simple theory holds only for weak drive powers when the transitions are governed by low-order mixing processes. In experiments, higher-order processes involving spurious modes are often significant, and indeed can help mediate intrinsic multi-photon transitions. As an example, in Fig.~\ref{fig:two_mode_floquet}(a) we focus on the experimental data for a range of drive frequencies between $\omega_d/2\pi = 10$ GHz and $\omega_d/2\pi =11.5$ GHz. The feature labeled \textbf{M} is identified as two dominant transitions $\ket{1_t}\rightarrow \ket{7_t}$ and $\ket{1_t}\rightarrow \ket{11_t}$, as verified from the Floquet simulation and a branch analysis. However, the two prominent features, \textbf{K} and \textbf{L} remain absent in the Floquet simulation data shown in Fig.~\ref{fig:two_mode_floquet}(b), indicating that these are mediated by spurious electromagnetic modes of the system. 

Indeed, we expect an inelastic scattering process involving the readout mode near $\omega_d/2\pi = 10.7$ GHz, as shown by the black dashed line in Fig.~\ref{fig:two_mode_floquet}(a). This is a six-wave mixing process where two drive photons are converted into an excitation in the readout resonator and three excitations in the transmon. We use the notation $1,0\rightarrow4,1$ to describe this process, with the two elements of the tuple representing the transmon and resonator excitations, respectively. Although the resonance condition agrees in the zero drive power limit, the experimental data show no agreement to this simple theory at higher drive powers. To explain the experimental data, we perform a Floquet simulation, with a larger system Hamiltonian that includes both the transmon and the readout mode; see Appendix~\ref{app:floquet_extrinsic_details} for details. 
Remarkably, the Floquet simulation captures both the shape and location of the resonance features, confirming that these transitions are inelastic processes mediated by the readout mode. 

From a branch analysis at lower drive powers, we identify both features \textbf{K} and \textbf{L} as inelastic scattering processes involving the readout mode, as shown in Fig.~\ref{fig:two_mode_floquet}(c). The feature \textbf{L} is confirmed to be the $1,0\rightarrow4,1$ process, while resonance \textbf{K} is identified to be $1,0\rightarrow7,1$. Note that these transitions may further couple to other multi-photon processes involving higher levels of the transmon, thus inheriting significant sensitivity to offset charge.

It is thus critical to simulate the effect of these electromagnetic modes in the relevant drive frequency range, when designing physical circuits.
In principle, a Floquet simulation with the system Hamiltonian that includes \textit{all the relevant electromagnetic modes} should accurately predict the experimental DUST spectrum.
However, this approach is not scalable, as the dimension of the Hilbert space grows exponentially with an increasing number of modes, and the computational cost of the Floquet simulation eventually becomes impractical. 
Building an efficient analysis framework that accounts for the Josephson potential along with its environment remains an open question.

%%=============================================================
%% TLS-ASSISTED INELASTIC SCATTERING
%%=============================================================

\section{TLS-assisted inelastic scattering}
\label{sec:inelastic_tls}

%%-------------------------------------------------------------
%%% Figure 8: Inelastic scattering processes (TLS)
%%-------------------------------------------------------------
\begin{figure}[t]
\includegraphics[width = 0.48\textwidth]{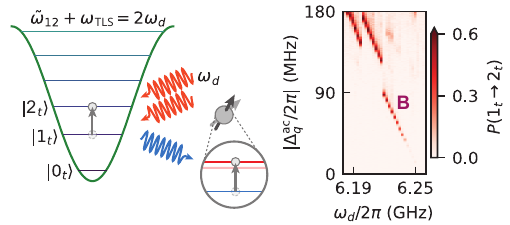}
\caption{Spurious transition caused by inelastic scattering assisted by a TLS with a switching frequency. (left) As the TLS switches between two frequencies, the combination of drive frequency and power at which the transition is activated also switches between two sets of values. (right) Measured population of the transmon state $\ket{2_t}$ under the drive. As the TLS switches its frequency twice during the timescale of the experiment, the resonance appears as a broken line.}
\label{fig6:TLS_heating} 
\end{figure}
%%-------------------------------------------------------------

In the previous section, we established that a spurious electromagnetic mode of the device or package may hybridize with the transmon and contribute to DUST.
Such processes are also mediated by other spurious degrees of freedom such as a TLS defect~\cite{Sank_2016_MIST}. Unlike the resonant exchange with a TLS near the qubit frequency, discussed earlier in Sec.~\ref{sec:tls_ac_stark_shift}, these transitions are activated by mixing processes involving a TLS far detuned from the qubit frequency. 
In Fig.~\ref{fig6:TLS_heating}, we focus on one such transition identified from the broadband spectroscopy data (labeled B in Fig.~\ref{fig:landscape}(b)). 

From the readout signal, we identify this transition as a single quantum excitation of the transmon from $\ket{1_t}$ to $\ket{2_t}$. 
This transition is consistent with a two-mode squeezing interaction between the transmon and an extrinsic mode around $8.17$ GHz. 
However, the broken line corresponding to the transition peak suggests that this extrinsic mode is switching between two well-defined frequencies spaced by $15$ MHz. 
This switching behavior makes the transition distinct from those discussed in Sec.~\ref{sec:geometric_inelastic} which arise from stable electromagnetic modes of the system.  To further characterize this transition, we monitor the single-shot readout data (See Appendix~\ref{app:tls_mpt_dynamics} for details) and observe a telegraphic ``on-off'' switching behavior. The timescale of switching of this resonance is several minutes, similar to some of the slow switching TLS we observe near the qubit frequency as shown in Fig.~\ref{fig:tls_spectrum}(c). We thus infer that this switching resonance is mediated by a strongly-coupled TLS, at a switching frequency $\omega_{\rm{TLS}}/2\pi = \{8.17, 8.185\}$ GHz.

Although we only show a four-wave-mixing process involving a TLS, in principle, they can also contribute to higher-order mixing processes.
Our results emphasize that TLS can mediate deleterious transitions even when these TLS are far detuned from the qubit frequency. 
Unlike the electromagnetic modes that can be accounted for in the device design process, the microscopic TLS bath is (as yet) unpredictable. 
Reducing the presence of TLS defects for a wide range of frequencies and eliminating their coupling to the transmon remains a crucial material challenge. 

On the other hand, our experiment establishes the time-resolved pump-probe spectroscopy as a novel technique for probing the TLS environment with a fixed frequency transmon. This technique probes the TLS environment through parametric four-wave mixing excitations. This approach, thus, not only eliminates the reliance on flux-tunable qubits for TLS spectroscopy, but also extends the accessible frequency range for such experiments. This technique paves the way for several research directions, including: (i) measurement of spectral density of TLS in the frequency range spanning up to several times the qubit frequency, and temporal fluctuations of these TLS, (ii) determination of the physical locations of these high-frequency TLS on the device through stress-strain tuning~\cite{bilmes_2021} and electric field tuning~\cite{Lisenfeld2019_TLS}, (iii) measurement of the dynamics of high-frequency TLS under background ionizing radiations~\cite{thorbeck_2023}, (iv) investigation of materials dependence~\cite{crowley_tantalum_2023} of the high-frequency TLS environment.

%%=============================================================
%% CONCLUSIONS
%%=============================================================
\section{Conclusion and Outlook}
\label{sec:conclusions}
The fidelity of microwave-driven operations in superconducting circuits is limited by the onset of drive-induced unwanted state transitions (DUST) as the drive power is increased. The coexistence of multiple distinct DUST mechanisms results in indiscriminate observations in experiments, obscuring the underlying causes and complicating mitigation strategies.
In this work, we systematically disentangle the mechanisms of DUST in a 3D transmon using time-resolved pump-probe spectroscopy. We identify and categorize three distinct mechanisms responsible for the unwanted transitions: In the first mechanism, \textit{A},  the drive-induced ac-Stark shift tunes the transmon levels into resonance with a spurious mode, such as a TLS, causing the transmon to exchange energy with the mode and relax to its thermal ground state.
In the second mechanism, \textit{B}, the drive activates an accidental multi-photon resonance within the transmon potential, exciting the transmon to non-computational states. 
Finally, in the third mechanism, \textit{C}, the drive activates an inelastic scattering process, in which it induces a transition in the transmon and scatters off at a different frequency, exciting a spurious mode coupled to the transmon.

With the pump-probe spectroscopy technique, we measure the drive-induced transition probability as a function of drive frequency and drive power and offer a strategy to parse these mechanisms.
When a transition does not explicitly depend on the drive frequency but only depends on the ac-Stark shift, we identify it as mechanism \textit{A}. If the transition appears in the Floquet simulation of the transmon Hamiltonian, we identify it as mechanism \textit{B}. All other transitions fall under mechanism \textit{C}. Most transitions from the third category are due to known RF modes in the physical device and are predictable through electromagnetic simulations. 

Our work suggests a set of layered mitigation strategies for DUST, each tailored to address a specific mechanism \textit{A}, \textit{B} or \textit{C}. The spurious modes near the qubit frequency, causing mechanism \textit{A}, are mainly comprised of TLS materials defects. Suppressing these requires improved materials and fabrication techniques~\cite{crowley_tantalum_2023, ganjam2024, bland2025}, reducing the TLS density and their coupling to the qubits. 
Mechanism \textit{B}, intrinsic multi-photon resonances, is captured by Floquet simulation of the driven Hamiltonian and can be avoided by engineering the drive frequency in a window free from these transitions. Importantly, drifting offset charge modifies the resonance conditions of these transitions, reducing the span of these windows. This problem can be addressed by either suppressing the offset charge sensitivity in hardware design~\cite{verney_2019, Wang2025}, or by stabilizing the offset charge via feedback~\cite{Fechant2025}. 

Finally, mechanism \textit{C} is caused by any spurious degrees of freedom, such as TLS defects or parasitic RF modes formed by the physical device layout. The latter is predictable and should be simulated during any device design process. Critically, our results show that the electromagnetic environment of the device, far beyond its immediate operating band, contributes to DUST processes. 
For example, the feature \textbf{N} in Fig.~\ref{fig:landscape}(c) is mediated by an RF mode at $28$ GHz. Therefore, mitigating these processes requires filtering and RF engineering for this entire range of frequencies to suppress undesired inelastic scattering processes. By extension, our findings also indicate that TLS do not need to be near-resonant with the qubit to participate in DUST mechanisms. Reducing TLS densities across wide frequency ranges is thus crucial to avoid mechanism \textit{C}.

The time-resolved pump-probe spectroscopy technique demonstrated here serves as a diagnostic tool and complements Floquet simulations and electromagnetic simulations. Through these simulations, many DUST mechanisms can be predicted in advance during the design phase. However, experimental validation using the spectroscopic method is necessary to detect unmodeled ``parasitic'' modes and validate the microwave hygiene.

The spectroscopy technique also establishes a foundation for investigating high-frequency TLS with a fixed frequency transmon, extending the accessible frequency range for such experiments.

Several challenges remain. Although Floquet simulations quantitatively predict experimentally observed resonance conditions, the relation between simulated hybridization parameters and measured transition probabilities remains qualitative. Establishing a quantitative connection is an important direction. Moreover, extending Floquet simulations to multiple modes is computationally costly and not scalable, motivating alternative approaches~\cite{Xiao2023, baskov2025exact}. Another key question is how to connect the impedance of the qubit environment, Re[$Z(\omega)$], to the inelastic scattering rates quantitatively~\cite{connolly_2025}. Finally, quasiparticle generation by strong drives and quasiparticle-assisted transitions~\cite{kishmar2025quasiparticle, chowdhury2025} represent important but unexplored contributors to DUST, warranting further investigation.

Together, our findings establish a systematic framework for identifying, modeling, and avoiding drive-induced unwanted state transitions in superconducting circuits, advancing high-fidelity operations in quantum information processing.

%%=============================================================
%% ACKNOWLEDGMENTS & AUTHOR CONTRIBUTIONS
%%=============================================================
\acknowledgments
The authors acknowledge Roman Baskov, Akshay Koottandavida and Alessandro Miano for insightful discussions. This research was sponsored by the Army Research Office (ARO) under grant nos. W911NF-23-1-0051, by the Air Force Office of Scientific Research (AFOSR) under grant FA9550-19-1-0399 and by the U.S. Department of Energy (DoE), Office of Science, National Quantum Information Science Research Centers, Co-design Center for Quantum Advantage (C2QA) under contract number DE-SC0012704. The views and conclusions contained in this document are those of the authors and should not be interpreted as representing the official policies, either expressed or implied, of the ARO, AFOSR, DoE or the US Government. The US Government is authorized to reproduce and distribute reprints for Government purposes notwithstanding any copyright notation herein. Fabrication facilities use was supported by the Yale Institute for Nanoscience and Quantum Engineering (YINQE) and the Yale University Cleanroom.
L.F. is a founder and shareholder of Quantum Circuits Inc. (QCI).

\section*{Data Availability}
All experimental datasets and analysis scripts used to generate the figures in this manuscript are publicly available in the Zenodo repository~\cite{dustdata}. 
This includes all raw and processed data required to reproduce the reported experimental results. 
Simulation results presented in this work can be reproduced using the open-source package \texttt{floquet}~\cite{floquet2024}.
Any additional data related to this manuscript are available from the authors upon reasonable request.

\section*{Author Contributions}
\textbf{W.D.:} Conceptualization (lead), Methodology (equal), Investigation (lead), Data curation (equal), Formal analysis (lead), Software (equal), Resources (equal), Writing-original draft (equal),
\textbf{S.H.:} Conceptualization (lead), Methodology (equal),  Investigation (equal), Data curation (equal),  Formal analysis (lead), Visualization (lead), Software (equal), Writing-original draft (lead), 
\textbf{D.K.W.:} Conceptualization (lead), Methodology (equal), Investigation (lead), Formal analysis (lead), Software (lead), Writing-original draft (equal),
\textbf{P.D.K.:} Conceptualization (equal), Methodology (supporting), Validation (supporting), Writing-original draft (supporting),
\textbf{T.C.:} Conceptualization (equal), Methodology (supporting), Validation (supporting),
\textbf{H.B.:} Methodology(supporting), Formal analysis (equal), Writing-original draft (supporting),
\textbf{S.S.:} Methodology (supporting), Formal analysis (supporting), Writing-original draft (supporting),
\textbf{V.R.J.:} Resources (supporting),
\textbf{A.Z.D.:} Resources (supporting), Writing-review \& editing (equal),
\textbf{P.D.P:} Software (supporting),
\textbf{J.V.:} Resources (supporting),
\textbf{X.X.:} Resources (supporting),
\textbf{L.F.:} Project administration (equal), Resources (equal), Supervision (equal), Writing-review \& editing (equal). 
\textbf{M.H.D.:} Conceptualization (equal), Funding acquisition (lead), Project administration (equal), Supervision (lead), Writing-review \& editing (equal).

%%=============================================================
%% END OF MAIN TEXT: BEGINNING OF APPENDIX
%%=============================================================
%%=============================================================
%% APPENDIX PREAMBLES
%%=============================================================
\appendix
\renewcommand{\thefigure}{A\arabic{figure}}
\renewcommand{\thetable}{A\arabic{table}}
\renewcommand{\theequation}{\thesection\arabic{equation}}
\setcounter{figure}{0}  
\setcounter{table}{0}  
\setcounter{equation}{0}  

%%=============================================================
%% Appendix A: Experimental Device
%%=============================================================
\section{Experimental Device}
\label{sec:device}

The experiment is performed on a fixed-frequency transmon qubit housed in a 3D rectangular waveguide cavity\cite{Paik2011_3D}. 
The fundamental mode (TE101) of the cavity serves as the readout mode.
The cavity has two ports coupled to external coaxial transmission lines. 
One port, which is used for reflection measurement, is strongly coupled, dominating the damping rate $\kappa_r/2\pi = 7.20$ MHz of the readout mode. 
The other port is weakly coupled to the cavity, but has a direct capacitive coupling to the transmon. 
We reduce the total attenuation of this line to strongly drive the transmon across a wide range of frequencies. 
However, since the port is only weakly coupled to the cavity, the reduced attenuation does not introduce significant photon shot noise. 
At the mixing-chamber stage of the dilution refrigerator, the drive line includes $20$~dB of attenuation. We verify that the attenuation is sufficient to preserve qubit coherence, while delivering adequate drive power for the spectroscopy experiment.

%%-------------------------------------------------------------
%% Figure A1: Readout signal
%%-------------------------------------------------------------
\begin{figure}[b]
\includegraphics[width = 0.48\textwidth]{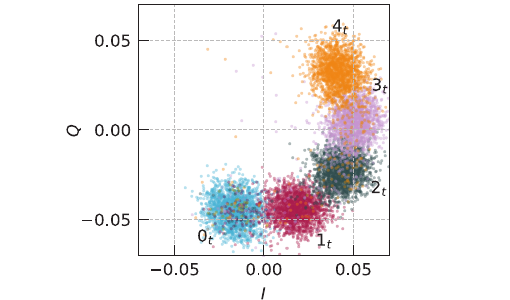}
    \caption{An example of a multi-state readout signal for the prepared transmon levels, $\ket{0_t}$, $\ket{1_t}$ , $\ket{2_t}$ , $\ket{3_t}$  and $\ket{4_t}$, constructed from $2000$ shots for each prepared states.  
    The transmon is prepared into the desired state by a sequence of selective pi pulses on appropriate transitions. The readout pulse is $500$ ns long and the reflected signal is integrated for $600$ ns to achieve a single-shot readout result. The ratio of the total dispersive shift to the readout resonator linewidth is chosen to be a small value, $\chi/\kappa\sim1/6$, to discriminate multiple transmon states with a single readout tone.
    }
\label{figA:ro_histogram}
\end{figure}
%%-------------------------------------------------------------

Our experimental device is the second \textit{qubit-cavity system} in the multiplexed readout setup in Ref.~\cite{joshi2025_lumped} and  is connected to the same broadband SNAIL parametric amplifier from the same reference.
The nonreciprocal components on the readout line make sure that the spectroscopy drive does not leak into the other cavity. 
A detailed discussion of the readout line setup is presented in Appendix D of \cite{joshi2025_lumped}. The average qubit relaxation time and the Hahn-echo time,  measured during this experiment, are $T_1\sim85$ $\mu$s and $T_2^E\sim135$ $\mu$s, respectively.
The measured total thermal population of the transmon excited states is $1.6\%$.

To measure the transmon state, we send a readout pulse of $500$ ns duration, and integrate the output signal for $600$ ns.
The readout scheme is optimized to discriminate multiple qubit levels in our two-tone spectroscopy experiment, resolving the final states after the state transition. The single shot readout signals for the prepared states, $\ket{0_t}$, $\ket{1_t}$ , $\ket{2_t}$ , $\ket{3_t}$  and $\ket{4_t}$  are shown in Fig.~\ref{figA:ro_histogram}.
Note that, in the readout calibration, we limit the readout power to ensure that the readout pulse itself does not activate unwanted state transitions. 
This precaution guarantees that any observed transitions arise solely from the stimulation drive, and are not artifacts of the readout process.

To set up a circuit model for the Floquet analysis, we first
experimentally characterize the transmon levels up to $\ket{5_t}$, and fit a simple cosine Josephson potential to the measured spectrum. 
The agreements between the measured levels and the fitted spectrum are summarized in Table.~\ref{tab:transmon_ladder}. This set of fitted circuit parameters is used in the Floquet simulation presented in the main text Sec.~\ref{sec:intrinsic_mpt} and Sec.~\ref{sec:floquet_sim}.
Note that the extracted model parameters differ when additional degrees of freedom, such as the readout mode, are explicitly included in the circuit model. We choose the appropriate model consistent with the scope of a given analysis.

%%-------------------------------------------------------------
%% Table A1: Transmon spectrum
%%-------------------------------------------------------------
\begin{table}[b]
    \centering
    \begin{tabular}{|c|c|c|c|}
        \hline 
        \makecell{Energy\\level} & \makecell{Measured\\spectra\\(MHz)}  & \makecell{Circuit model\\(transmon only)\\(MHz)} & \makecell{Circuit model\\(transmon \\ and RO)\\(MHz)}\\
        \hline 
        $E_{|0_t\rangle}/h$ & $0$ & $0$ & $0$\\
        $E_{|1_t\rangle}/h$ & $4528.52$ & $4531.06$ & $4528.05$\\
        $E_{|2_t\rangle}/h$ & $8872.74$ & $8875.30$ &$8871.66$\\
        $E_{|3_t\rangle}/h$ & $13016.8$ & $13018.4$ &$13016.7$\\
        $E_{|4_t\rangle}/h$ & $16939.9 \pm 0.2$ & $16941.7$ &$16944.8$\\
        $E_{|5_t\rangle}/h$ & $20613 \pm 2$ & $20619.1$ & $20631.9$\\
        \hline
    \end{tabular}
    \caption{Measured transmon spectra and the fitted spectra from two minimal circuit models. The levels $E_{|4_t\rangle}$ and $E_{|5_t\rangle}$ exhibit frequency dispersion.
    The first circuit model fits the transmon spectrum to a cosine potential alone with $E_{J}/h=16.2856$ GHz, $E_{C}/h=0.17013$ GHz. The second circuit model includes the readout (RO) resonator mode and assumes a linear charge-charge coupling between the transmon dipole and the readout resonator. The corresponding fit parameters are $E_{J}/h=16.40$ GHz, $E_{C}/h=0.1695$ GHz, $g/h=0.153$ GHz, $\omega_{r}/2\pi=9.029$ GHz.}
    \label{tab:transmon_ladder}
\end{table}
%%-------------------------------------------------------------

%%=============================================================
%% Appendix B: Time-resolved two-tone spectroscopy
%% from the ground and excited states
%%=============================================================
\section{Time-resolved two-tone spectroscopy from the ground and excited states}
\label{sec:landscape_g_and_e}

In Section~\ref{sec:dust_spectroscopy} of the main text, we have introduced the time-resolved two-tone spectroscopy experiment to systematically investigate the occurrence of DUST, as a function of drive frequency and drive power. 
In this appendix, we present additional experimental details, and analyze the results of the observed DUST landscape, both from $\ket{0_t}$ (ground state) and $\ket{1_t}$ (first excited state) of the transmon.

\subsection{Room temperature hardware and power calibration}
The stimulation drive is synthesized by applying a pair of baseband DAC channels as IQ inputs to a Marki MMIQ-0218LXPC mixer, operated in zero-sideband mode. The drive frequency is swept by tuning the local oscillator (LO) frequency. At each $\omega_d$, we first perform LO leakage cancellation, and then calibrate the ratio between the induced $\Delta_q^{\rm ac}$ and the applied drive power (controlled by the squared DAC amplitude).
The drive power at each frequency is subsequently scaled, so that the corresponding induced ac-Stark shift values are consistent across the entire frequency sweep, spanning a range from $\Delta_q^{\rm ac}=0$ to $\Delta_q^{\rm ac} = \alpha_q$. 
This is equivalent to sweeping the square of the applied drive strength, $\xi^2$, from $0$ to $2$, as introduced in Appendix~\ref{sec:xi}.

To sweep across frequencies, we adopt different configurations of room-temperature microwave components, such as amplifiers and band-pass filters, to achieve the desired power levels within each frequency window. This window-specific sharp filtering is essential to suppress spurious tones and harmonics, generated by the active components in the room-temperature setup.
For each configuration, we independently examine the generated drive tone in a spectrum analyzer and confirm that all active components are operated in their linear regime and that no spurious tone is generated over the relevant range of drive powers. This ensures (i) the ac-Stark shift calibration, which is performed at low drive strength, remains valid across the full sweep range at each frequency, and (ii) the pump being sent for the spectroscopy is a monochromatic drive. 

As we approach the readout resonator frequency (marked by the violet line), the transfer function of the drive, seen by the transmon, is dominated by the frequency response of the readout resonator. 
Due to the dispersive interaction, the resonator frequency depends on the qubit state, making the transfer function itself dependent on the qubit state.
Thus, near the resonator frequency, the Stark shift is calibrated separately for $\ket{0_t}$ and $\ket{1_t}$ initialization to account for this difference in the transfer function.

\subsection{Measured DUST probabilities}
Although we can resolve up to $\ket{4_t}$ using our readout, for the spectroscopy experiment producing DUST landscape, we set the readout threshold to distinguish states $\ket{0_t}$, $\ket{1_t}$ and $\ket{2_t}$ from all other states. The reason for this choice will be explained in the following subsection. After the readout, we wait for $~1$ ms for the system to relax into the thermal ground state. 

\begin{figure}[t]
\includegraphics[width = 0.48\textwidth]{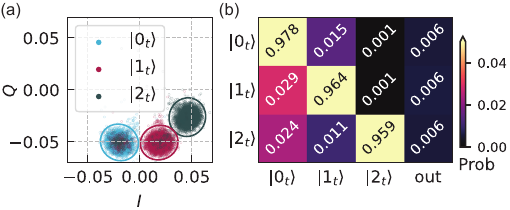}
    \caption{Mapping the readout signal to inferred transmon states, $\ket{0_t}$, $\ket{1_t}$, $\ket{2_t}$,  and \textit{outliers} using thresholds for the DUST spectroscopy experiment. 
    (a) Readout signal of prepared $\ket{0_t}$, $\ket{1_t}$, $\ket{2_t}$ states are constructed from $10000$ shots each. The mean and the variances of these distributions are then computed from a Gaussian mixture model. The ellipses on the quadrature space represent the $3\sigma$ intervals of the respective gaussian fits to the readout signals. These thresholds are applied for the state assignment in the spectroscopy experiment. 
    (b) Confusion matrix representing the probabilities of transmon state assignment. Each cell shows the probability of assigning a particular state given a prepared Fock state. Rows correspond to the prepared states, while columns correspond to the assigned states: $\ket{0_t}$, $\ket{1_t}$, $\ket{2_t}$, and outliers (labeled ``out''). Note that the state preparation in this characterization was performed without pre-selection readout, resulting in additional errors due to residual thermal population.
    }
\label{figA:ro_assignment}
\end{figure}

The drive-induced state transition data from $\ket{0_t}$ and $\ket{1_t}$ are acquired in an interleaved fashion to suppress the effects of slow experimental drifts, ensuring a consistent comparison between the two datasets across the entire sweep. For each drive frequency, the entire experiment, including mixer calibration, power calibration, and power sweep, takes approximately $5$ minutes. 
Also, note that the two datasets for the DUST landscape, above and below the qubit frequency, were taken in two successive cooldowns. Because the dataset was acquired over several days and required multiple adjustments to the room-temperature filter configurations, we periodically recalibrated the qubit $\pi$ pulse, the readout pulse, and the threshold for state assignment to account for hardware variations and slow system drift. The readout signals and the corresponding confusion matrix obtained after one such calibration are shown in Fig.~\ref{figA:ro_assignment}.

\begin{figure*}
    \centering
    \includegraphics[width=0.96\textwidth]{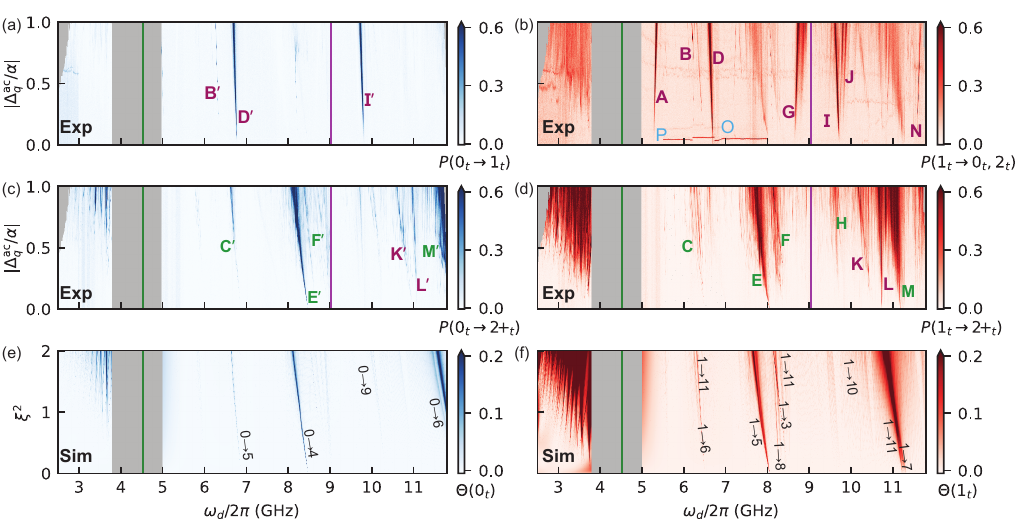}
    \caption{Landscape of drive-induced state transition: experimental data and Floquet simulation result. (a-d) Experimentally measured probability of transition out of the initially prepared states. Panels (a) and (b) show the single-quantum transitions out of $0_t$ and $1_t$ respectively. Panels (c) and (d) show the transitions into excited states states higher than $|2_t\rangle$. We have labeled transitions of different mechanisms in fonts of different colors. Note that the data for drive frequencies above and below the transmon frequency are taken in two different cool-downs. 
    (e-f) The hybridization parameter $\Theta(0_t)$ and $\Theta(1_t)$, respectively, obtained from Floquet simulation. We have labeled the identified transitions in the simulated plot. }
\label{fig:landscape_floquet_g_e}
\end{figure*}

Fig.~\ref{fig:landscape_floquet_g_e}(a-d) presents the measured transitions in the time-resolved two-tone spectroscopy experiment when the transmon is prepared in $\ket{0_t}$ (left column), and $\ket{1_t}$ (right column). In contrast to the main text Fig.~\ref{fig:landscape}, here we separate the final state populations into two sub-classes. In the top row, we plot the transitions that causes single quanta excitation (i.e. $\ket{0_t}\rightarrow \ket{1_t}$ and $\ket{1_t}\rightarrow\ket{2_t}$)  or relaxation ($\ket{1_t}\rightarrow\ket{0_t}$) of the transmon. And in the second row we plot the transition probabilities to all other states. This segregation helps us to compare the measured transitions with the Floquet simulation.

Within the drive range used in this experiment, mechanism \textit{B} can only excite transmon states above $\ket{2_t}$. Intrinsic multi-photon processes that connect $\ket{0_t}$ to $\ket{1_t}$ or $\ket{1_t}$ to $\ket{2_t}$ lie in the subharmonic regime~\cite{xiaFastSuperconductingQubit2023}, which falls outside the frequency range covered in this plot. 
As a result, the top panel contains no transitions arising from mechanism \textit{B}. 
All visible transitions are instead attributed either to mechanism \textit{A}, if they appear as quasi-horizontal lines, or to mechanism \textit{C}, if they display a slope in frequency-power space.
Note that, while transitions to higher transmon levels should not be present, we observe faint residual signatures of them in Fig.~\ref{fig:landscape_floquet_g_e}(b). These are artifacts of imperfect state discrimination between $\ket{2_t}$ and higher excited states during readout.

%%-------------------------------------------------------------
%% Figure A2: Landscape of drive-induced state transitions
%%-------------------------------------------------------------
%%-------------------------------------------------------------

Comparing the left (a,c) and right (b,d) columns in Fig.~\ref{fig:landscape_floquet_g_e}, we observe that several features appear in pairs and exist in both cases, corresponding to the two different transmon initializations, $\ket{0_t}$ and $\ket{1_t}$. 
Moreover, transition features in the $\ket{0_t}$ dataset occur at consistently higher drive frequencies than their $\ket{1_t}$ counterparts.
This shift arises from the negative anharmonicity of the transmon, which lowers the resonance conditions for processes involving $\ket{1_t}$ than those involving $\ket{0_t}$. 
We label these pairs of transitions with the same roman letters, with and without an apostrophe, such as \textbf{B\ensuremath{'}} and \textbf{B}.
A notable exception is a class of transitions, such as \textbf{A}, \textbf{G}, \textbf{J}, and \textbf{N}, that exhibit an opposite (positive) slope, and appear exclusively under $\ket{1_t}$ initialization. 
These are $\ket{1_t} \rightarrow \ket{0_t}$ decay processes caused by inelastic scattering, which have no counterpart for the ground state. 

Finally, for comparison, we show in panels (e,f) the results from the transmon-only Floquet simulations (see Appendix~\ref{sec:floquet_sim}), which compute the hybridization parameter $\Theta(0_t)$ and $\Theta(1_t)$ as a function of drive frequency and power. The transitions in Fig.~\ref{fig:landscape_floquet_g_e}, labeled with green letters, are reproduced by the Floquet simulation, aligning with the observed resonance conditions. However, some multi-quanta excitations, such as \textbf{K}, \textbf{L} and \textbf{K\ensuremath{'}}, \textbf{L\ensuremath{'}} are not captured in the Floquet simulation, confirming their origin in inelastic scattering processes. These transitions are discussed in Appendix~\ref{app:floquet_extrinsic_details}.

%%=============================================================
%% Appendix C: Derivation of the dimensionless drive
%% strength xi in the perturbative picture
%%=============================================================
\begin{section}{Derivation of the dimensionless drive strength $\xi$ in the perturbative picture}
\label{sec:xi}

In this appendix, we derive the dimensionless measure of the drive strength $\xi$ defined in Eq.~\eqref{eq:xi}. 
This quantity bridges the Floquet analysis (elaborated in Appendix \ref{sec:floquet_sim}) and the perturbative theory of driven nonlinear oscillators \cite{Xiao2023}. 
We begin with the following Hamiltonian of a driven transmon:
\begin{align}
\label{eq:Hamitlonian_appendix}
H(t) = 4E_{C}(\hat{n}-n_{g})^2 - E_{J}\cos(\hat{\varphi}) + E_d \hat{n} \cos(\omega_d t).
\end{align}
where we consider the charge drive as a classical continuous wave (CW) drive at frequency $\omega_d$. 

For a transmon satisfying $E_C \ll E_J$, we approximately express the Hamiltonian in terms of the creation and annihilation operators~\cite{blais_2021}:
\begin{align}
\hat{H}(t) &\approx (\sqrt{8E_CE_J}-E_{C})\hat{b}^{\dagger} \hat{b} - \frac{E_C}{12}\left(\hat{b}^{\dagger} + \hat{b} \right)^4 \\ \nonumber  &- i n_{\ZPF} E_d (\hat{b}-\hat{b}^{\dagger}) \cos(\omega_d t) +E_{C}\hat{b}^{\dagger}\hat{b},
\end{align}
where we have kept the leading order nonlinearity of the transmon and extracted out the frequency renormalization resulting from normal ordering the quartic term. Thus, here we approximate the transmon as a Duffing oscillator, with anharmonicity $\alpha_q = -E_C/\hbar$. 

We account for the linear drive by a time-dependent frame transformation~\cite{Xiao2023}
$\hat{U}_d=e^{(\beta_{\rm{lin}}^{*}\hat{b} - \beta_{\rm{lin}} \hat{b}^{\dagger})}$, and choose $\beta_{\rm{lin}}$ to cancel the linear term in $\hat{H}' = \hat{U}^{\dagger}_d\hat{H}\hat{U}_d-i\hat{U}^{\dagger}_d\dot{\hat{U}}_d$. 
We find
\begin{equation}
\label{eq:beta_lin}
\beta_{\rm{lin}} = \frac{i n_{\ZPF} E_d }{2 \hbar} \left( 
\frac{e^{-i \omega_d t}}{\omega_d - \omega_q} - \frac{e^{i \omega_d t}}{\omega_d + \omega_q} \right),
\end{equation}
where $\hbar\omega_q = \sqrt{8E_CE_J}-E_{C}$, valid under the assumption $|\omega_d - \omega_q| \gg |\alpha|$. 
The resulting Hamiltonian in the displaced frame is: 
\begin{align}
\label{eq:app_H_disp}
\hat{H}'(t)/\hbar
&= \omega_q \hat{b}^{\dagger} \hat{b} + \frac{\alpha_q}{12}\left(\hat{b}^{\dagger} + \hat{b} + \xi \sin[\omega_d t]\right)^4 \\ \nonumber 
&\quad- \alpha_{q}(\hat{b}^{\dagger}+\beta_{\text{lin}}^{*})(\hat{b}+\beta_{\text{lin}}),
\end{align}
where $\beta_{\rm{lin}} + \beta_{\rm{lin}}^{*}=\xi \sin(\omega_d t)$ and the term in the second line of Eq.~\eqref{eq:app_H_disp} disappears upon normal ordering the quartic term.
The effect of the drive on a weakly nonlinear oscillator is thus captured by a time-dependent displacement of the potential, where
\begin{equation}
\label{eq:app_xi}
\xi = \frac{2 n_{\ZPF} \omega_d}{\omega_d^2 - \omega_q^2} \frac{E_d}{\hbar}.
\end{equation}
We emphasize that in the Floquet simulations, no quartic expansion of the potential is made, and we retain the full cosine nonlinearity. 

Assuming that the drive does not activate any spurious processes, the Hamiltonian reduces to 
\begin{equation}
\hat{H}'(t)/\hbar \approx \left(\omega_{q} + \xi^2 \alpha_q/2\right)\hat{b}^{\dagger} \hat{b} + \frac{\alpha_q}{2} \hat{b}^{\dagger} \hat{b}^{\dagger} \hat{b} \hat{b},
\end{equation}
under the rotating-wave approximation. We have thus additionally derived the relationship between $\xi$ and the induced ac-Stark shift on the transmon, $\Delta_{q}^{\rm{ac}}=\xi^2\alpha_{q}/2$.
\end{section}

%%=============================================================
%% Appendix D: Floquet simulation for predicting
%% intrinsic multi-quanta resonances
%%=============================================================

\begin{section}{Floquet simulation for predicting intrinsic multi-quanta resonances}
\label{sec:floquet_sim}

In this appendix, we discuss the Floquet simulation techniques generating the Floquet overlap and branch analysis plots presented in the main text. 
Our implementation of both Floquet simulation techniques is available in the open-source \texttt{floquet} package \citep{floquet2024}. 

To distinguish mechanism \textit{B}, we map out the landscape of intrinsic multi-photon transitions as a function of drive frequency and power for a transmon, which is captured by the Hamiltonian as Eq.~\eqref{eq:Hamitlonian_appendix}. 
Note that previous works investigating the phenomenon of measurement-induced state transitions model the readout-resonator photons as a stiff, classical drive at the readout frequency \cite{Shillito2022, Cohen2023, Dumas_2024_Ionization, Xiao2023}, resulting in the same Hamiltonian as Eq.~\eqref{eq:Hamitlonian_appendix}. 
That modeling is only appropriate provided that quantum fluctuations of the resonator and dynamics due to ramping of the drive can be ignored. 

In principle, all intrinsic multi-photon resonances associated with Eq.~\eqref{eq:Hamitlonian_appendix} can be systematically predicted by extracting the Floquet modes. 
By appropriate analysis of the Floquet modes (described in detail below) we produce both the ``overlap'' plots \cite{Xiao2023} that provide a birds-eye view of DUST as well as the ``branch-analysis'' plots \cite{Shillito2022, Cohen2023, Dumas_2024_Ionization} to assign transmon-eigenstate indices to Floquet modes.  

We first diagonalize the undriven transmon Hamiltonian in \texttt{scqubits} \cite{scqubits1,*scqubits2}, and express Eq.~\eqref{eq:Hamitlonian_appendix} in this new basis:
\begin{align}
\label{eq:transmon_diag}
H=\sum_{j}E_{j}|j_t\rangle\langle j_t| + E_d \cos(\omega_{d} t) \sum_{ij}n_{ij}|i_t\rangle\langle j_t|,
\end{align}
where $|j_{t}\rangle$ is the $j^{\text{th}}$ eigenstate of the transmon and $n_{ij}=\langle i_t|\hat n|j_t\rangle$. 
We then extract from \eqref{eq:transmon_diag} the Floquet modes $\{ \ket{\tilde{j}_t(\xi, \omega_d)} \}$ expressed in the basis of the undriven transmon eigenstates $\{\ket{j_t}\}$ \cite{floquet2024, qutip1, *qutip2}. 
We have written explicitly the dependence of the Floquet modes on the drive frequency $\omega_{d}$ and phase displacement $\xi$ (c.f. Eq.~\eqref{eq:app_xi}). 
As the junction response depends on the detuning from resonance, we express the drive strength in terms of $\xi$, as opposed to $E_{d}$, to compare results across different $\omega_{d}$. 
To compare with the experimental results, we perform these Floquet simulations over the same range of drive frequencies as in Fig.~\ref{fig:landscape}. 
We additionally scan over drive amplitudes $0\leq\xi^2\leq2$, such that the induced ac-Stark shift (energy difference between $\ket{\tilde{0}_t}$ and $\ket{\tilde{1}_t}$) ranges up to $\alpha_q$, consistent with the experiment. 
In our simulations the drive strength interval is chosen to be $\delta\xi^2=0.005$, matching that of the experimental scan, which corresponds to a step size in Stark shift of approximately 460 kHz.
In all cases, we truncate to 25 states in the transmon Hilbert space.
We discuss further below how we assign a transmon-eigenstate label $j$ to the Floquet mode $\{ \ket{\tilde{j}_t(\xi, \omega_d)} \}$.

\begin{subsection}{Hybridization parameter method}
The off-resonant microwave drive primarily induces two effects. The first is the aforementioned multi-photon resonances. The second and less interesting effect is the ac-stark shift, which causes the overlap of the Floquet mode with the associated bare transmon eigenstate to decrease with increasing drive strength (even in the absence of DUST). To disentangle these effects, we utilize the ``ideal-displaced state'' \cite{Xiao2023, Kurilovich2025} which is constructed by fitting a 2D polynomial to the Floquet modes in drive amplitude and frequency space. 

For a given bare transmon state $\ket{j_{t}}$ for which we want to track DUST, we first identify the corresponding Floquet mode $|\tilde{j}_{t}(\xi,\omega_{d})\rangle$ by taking overlaps of the modes with $\ket{j_{t}}$. This is done at all computed pairs of $(\xi, \omega_{d})$. We then construct the ideal-displaced state $|\bar{j}_{t}(\xi,\omega_{d})\rangle$ in the basis of the bare transmon states as a low-order polynomial in $\xi,\omega_{d}$ \cite{Xiao2023, floquet2024}
\begin{align}
\label{eq:app_displaced_state}
\langle i_{t}|\bar{j}_{t}(\xi,\omega_{d})\rangle=\sum_{k,\ell}C_{ijk\ell}\xi^{k}\omega_{d}^{\ell}.
\end{align}
We optimize the coefficients $C_{ijk\ell}$ such that this state closely approximates $|\tilde{j}_{t}(\xi,\omega_{d})\rangle$. Importantly, we institute a cutoff (typically taken to be 0.8) whereby only Floquet modes with overlap with the bare state above this cutoff are included in the fit. As such, the state $|\bar{j}_{t}(\xi,\omega_{d})\rangle$ should capture the effects of ac-Stark shift, with any deviation indicating a resonance. In practice, we often need to iterate this procedure, as the overlap of the Floquet mode with the bare state $\ket{j_{t}}$ eventually falls below the cutoff for large enough ac-Stark shift (causing the assignment of labels to Floquet modes to fail), even in the absence of a resonance. In this case, we partition the drive amplitudes, and use the fitted displaced state from the previous partition as the bare state \cite{Xiao2023, Kurilovich2025, floquet2024}.

From the definition of the ideal-displaced state $|\bar{j}_{t}(\xi,\omega_{d})\rangle$, any deviation of the Floquet mode $|\tilde{j}_{t}(\xi,\omega_{d})\rangle$ from this state indicates a resonance. As such, we introduce the measure
\begin{align}
\Theta(j_t) = 1 - |\langle \tilde{j}_{t}(\xi,\omega_d)|\bar{j}_t(\xi,\omega_d)\rangle|^2,
\end{align}
which quantifies the degree of hybridization of the Floquet mode with a non-computational state. This hybridization remains nearly zero in the absence of resonances. When the drive condition satisfies a multi-quanta resonance, the hybridization $\Theta(j_t)$ increases sharply. The hybridization parameter is evaluated independently at each drive strength and frequency, without relying on continuity tracking between adjacent parameter points. Consequently, this Floquet analysis method is robust to the choice of sampling interval in the parameter space.

Notably, this technique does not immediately provide information on {\it which} state(s) are involved in a resonance. To rectify this, we  perform the displaced-state fit for non-computational states, and pair up locations where we see both $\Theta(j_{t})$ and $\Theta(\ell_{t})$ simultaneously deviate from zero. However,this technique is often numerically unstable due to the enhanced Stark shifts experienced by higher-lying states. In addition, such a tracking procedure becomes confused by multiple independent processes occurring simultaneously, and also by processes involving more than two states. Therefore, to label the states, we turn to the alternative technique: branch-analysis.

\end{subsection}

\begin{subsection}{Branch analysis}
\label{app:branch_analysis}

The hybridization method introduced above provides a straightforward means of identifying locations in $(\xi, \omega_{d})$ space where multi-quanta resonances occur. However, this technique does not immediately specify which non-computational state(s) participate in the resonance. We thus also implement the so-called {\it branch-analysis} technique introduced in Refs.~\cite{Shillito2022, Cohen2023, Dumas_2024_Ionization}, which allows for state tracking and labeling of the transitions. 

For a given $\omega_{d}$, we track the Floquet modes as a function of drive amplitude. 
We assign state labels based on overlaps with the Floquet modes at the previous drive amplitude, beginning with the undriven transmon eigenstates. Having thus collected a set of Floquet modes $b_{i}(\omega_{d})=\{|\tilde{i}_{t}(\xi, \omega_{d})\rangle\}$ for each state in our Hilbert space, we compute $N_{i}(\xi, \omega_{d})=\sum_{j}j\langle j_t|\tilde{i}_t(\xi, \omega_{d})\rangle$, weighting each Floquet mode by the bare transmon quanta \cite{Shillito2022, Cohen2023, Dumas_2024_Ionization}. We identify the resonances and the associated states by observing the {\it branch swaps}, see Fig.~\ref{fig:intrinsic_resonance}(c,d).

\end{subsection}

\end{section}

%%=============================================================
%% Appendix D: Floquet simulation for predicting
%% intrinsic multi-quanta resonances
%%=============================================================

\section{Analysis of multi-quanta resonances involving extrinsic modes}
\label{app:floquet_extrinsic_details}
%%-------------------------------------------------------------
%% Fig A3: Lumped element mode
%%-------------------------------------------------------------
\begin{figure}[t]
\includegraphics[width = 0.48 \textwidth]{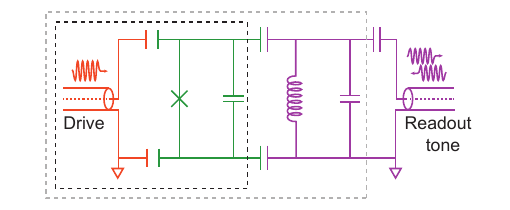}
\caption[Lumped element circuit]{Lumped element model for an ideal transmon-resonator system for dispersive readout scheme. For the transmon-only Floquet simulation (explained in Appendix~\ref{sec:floquet_sim}) we fit the experimentally observed spectrum to the circuit shown in the black dashed box to obtain the Hamiltonian parameters. The Floquet simulation
in Appendix~\ref{app:floquet_extrinsic_details} includes the readout resonator mode as shown by the dashed gray box.}
\label{figA:circuit_schematic} 
\end{figure}
%%-------------------------------------------------------------

Sections \ref{sec:geometric_inelastic} and \ref{sec:extrinsic_multi_photon} examined transitions mediated by spurious electromagnetic modes. Specifically, the transmon's unavoidable coupling to these modes leads to resonances corresponding to inelastic scattering of drive photons. As depicted in Figure~\ref{fig:app_inelastic_with_readout}, some of these transitions are predicted by incorporating an additional mode, such as the readout resonator, into Floquet simulations. 
This appendix extends the Floquet analysis from Appendix~\ref{sec:floquet_sim} to include coupling with spurious modes, particularly the readout resonator. Table~\ref{tab:inelastic_transitions} summarizes both intrinsic transmon transitions and those mediated by the readout resonator, in the frequency range shown in Fig.~\ref{fig:two_mode_floquet}, demonstrating agreement between experimental data and numerical simulations.

\subsection{The coupled transmon-resonator system}
\label{app:system}

Figure~\ref{fig:motivation}(d) highlights several spurious electromagnetic modes of the system that couple to the qubit. These electromagnetic modes are modeled as bosonic degrees of freedom capacitively coupled to the transmon. To manage computational complexity, we analyze one parasitic mode at a time. Here, as an example,  we focus on the readout mode, as the frequency of this mode and its coupling to the transmon are precisely obtained from the experiments.

The combined Hamiltonian of the transmon, coupled to the readout resonator is given by,
\begin{align}
\label{eq:app_coupled_Hamiltonian}
H &= 4E_{C}(\hat{n}-n_{g})^2 - E_{J}\cos(\hat{\varphi}) +\omega_{a}\hat{a}^{\dagger}\hat{a} \\ \nonumber 
&\quad  - ig \hat{n}(\hat{a}-\hat{a}^{\dagger})
+ E_d \hat{n} \cos(\omega_d t).
\end{align}
where $\hat{a}$ is the bosonic annihilation operator for the readout mode, $\omega_a$ its frequency, and $g$ is the coupling strength with the transmon. Same as before, $E_d$ and $\omega_d$ denote the drive strength and drive frequency, respectively. We use the parameters given in Table \ref{tab:transmon_ladder} obtained from fitting the experimental spectrum.

In our simulations, we truncate the joint transmon-readout mode system to $25 \times 5$ states, consistent with the number of transmon levels  in Appendix \ref{sec:floquet_sim}. We justify the truncation of the readout mode by confirming that the identified transitions remain unchanged when the Hilbert space is expanded.

\subsection{State labeling}\label{app:state_label}
Following the approach in Appendix \ref{sec:floquet_sim}, we begin by diagonalizing the undriven Hamiltonian \eqref{eq:app_coupled_Hamiltonian} in \texttt{scqubits} \cite{scqubits1,*scqubits2} and expressing Eq.~\eqref{eq:app_coupled_Hamiltonian} in this new basis,
\begin{align}
\label{eq:app_coupled_diag_Hamiltonian}
H&=\sum_{j}E_{j}\dyad{\overline{j}} + E_d \cos(\omega_{d} t) \sum_{j,k}n_{jk}\dyad{\overline{j}}{\overline{k}},
\end{align}
where $\ket{\overline{j}}$ indicates a dressed eigenstate in the coupled Hilbert space of the transmon and the resonator. It is usually possible to identify these dressed eigenstates with bare labels, i.e. $\ket{\overline{j}}=\ket{\overline{i_{t},k_{r}}}$, by finding the bare-product state $\ket{i_t, k_{r}}=\ket{i_t} \otimes \ket{k_{r}}$ with largest overlap. We institute a threshold (here, 0.9) below which bare labels are not assigned to a dressed state. Such cases typically arise due to strong hybridization between bare states, particularly higher-energy ones. 
As an example, the bare states $\ket{6_t,1_r}$ and $\ket{9_t,0_r}$ are nearly degenerate for the parameters of our system. The bilinear coupling between the transmon and the resonator breaks this degeneracy, hybridizing these two levels. Therefore, we cannot assign bare labels to the corresponding eigenstates $\ket{\overline{21}}$ and $\ket{\overline{22}}$, which are approximately even and odd superpositions of the two bare states.

%%-------------------------------------------------------------
%% Table A2: Examples of transitions K, L, K', L'
%%-------------------------------------------------------------
\begin{table}[t]
    \centering
    \begin{tabular}{|c|c|c|c|}
        \hline 
        \makecell{Label} & \makecell{Transition}  & \makecell{Frequency \\ $\omega_d/2\pi$} & \makecell{Resonance \\ condition} \\
        \hline
        \rule{0pt}{\normalbaselineskip} 
        $\textbf{K}$ & $\ket{\overline{1_t, 0_r}} \leftrightarrow \ket{\overline{7_t,1_r}}$ & 10.57 GHz & \makecell{$\omega_{1,7} + \omega_r = 3 \omega_d$} \\[3pt]
        $\textbf{L}$ & $\ket{\overline{1_t, 0_r}} \leftrightarrow \ket{\overline{4_t,1_r}}$ & 10.74 GHz & \makecell{$\omega_{1,4} + \omega_r = 2 \omega_d$} \\[3pt]
        \multirow{2}{*}{$\textbf{L}'$ \bigg\{}  & $\ket{\overline{0_t, 0_r}} \leftrightarrow \ket{\overline{3_t,1_r}}$ & 11.04 GHz & \makecell{$\omega_{0,3} + \omega_r = 2 \omega_d$} \\[3pt]
        & $\ket{\overline{0_t, 0_r}} \leftrightarrow \ket{\overline{22}}$ & 11.04 GHz & \makecell{$\omega_{0,9} = 3 \omega_d$} \\[3pt]
        $\textbf{K}'$ & $\ket{\overline{0_t, 0_r}} \leftrightarrow \ket{\overline{21}}$ & 11.03 GHz & \makecell{$\omega_{0,6} + \omega_r = 3 \omega_d$} \\[3pt]
        \hline
    \end{tabular}
    \caption{Examples of transitions and their corresponding resonance conditions in the joint transmon-resonator system for drive frequencies $\omega_d/2\pi \in [10.2, 11.4]$ GHz. Transitions are labeled based on their qualitative resemblance to experimental results shown in Figure~\ref{fig:landscape_floquet_g_e}. Here, $\omega_r$ and $\omega_{i,j}$ represent the readout mode frequency and the transition frequency between the transmon levels $\ket{i_t}$ and $\ket{j_t}$, respectively. Frequencies and resonance conditions are reported at zero-drive strength. 
    The resonance conditions in terms of the bare states for the final two rows  are randomly assigned, as the transitions involve the highly-hybridized states $\ket{\overline{21}},\ket{\overline{22}}$. 
    }
    \label{tab:inelastic_transitions}
\end{table}
%%-------------------------------------------------------------

Having identified the dressed states of interest, we now apply the hybridization parameter method and branch analysis technique as outlined in Appendix \ref{sec:floquet_sim}. There is no change to the analysis, aside from the difficulty in certain cases of identifying certain dressed states with bare labels. 

%%-------------------------------------------------------------
%% Figure A4: Floquet analysis of an inelastic scattering 
%% involving the readout mode
%%-------------------------------------------------------------
\begin{figure}[t]
    \centering
    \includegraphics[width=0.48\textwidth]{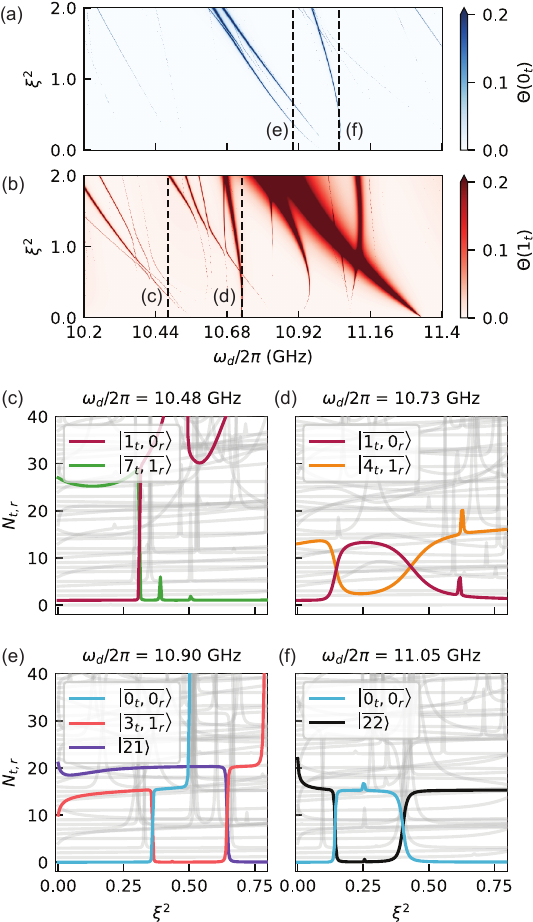}
    \caption{Floquet analysis of an inelastic scattering involving the readout mode, for a fixed gate charge $n_g=0.0$. Panels (a) and (b) depict the transition landscape in terms of the hybridization parameters $\Theta(0_t)$ and $\Theta(1_t)$, respectively, within the range $\omega_d/2\pi \in [10.2, 11.4]$ GHz. Panels (c)–(f) present branch analyses at specific frequencies, indicated by dashed lines in (a) and (b). The branches are identified with an undriven dressed state. Whenever possible, these undriven dressed state are labelled $\ket{\overline{i_t, k_r}}$ corresponding to the bare product state with which they have the most overlap. If they do not have a large overlap (>0.9) with any bare state, we assume the states are hybridized and label them with their dressed index, e.g. $\ket{\overline{21}}$ or $\ket{\overline{22}}$.}
    \label{fig:app_inelastic_with_readout}
\end{figure}
%%-------------------------------------------------------------

\subsection{Example of unwanted transitions assisted by the readout mode}
\label{app:inelastic_with_readout}
The readout resonator, a paradigmatic spurious mode, is necessarily coupled to the qubit and mediates several experimentally observed transitions. Figure~\ref{fig:app_inelastic_with_readout} illustrates the transition landscape and branch analysis at selected frequencies, derived from a Floquet analysis of the coupled transmon-resonator system. Table~\ref{tab:inelastic_transitions} summarizes some key transitions and links them to the experimentally observed transitions in Figure~\ref{fig:landscape_floquet_g_e}.

Similar to intrinsic transitions, these inelastic transitions occur in pairs for $\ket{\overline{1_t,0_{r}}}$ and $\ket{\overline{0_t,0_{r}}}$. For instance, transitions \textbf{K} and \textbf{K}$\mathbf{'}$ form such a pair, corresponding to the transitions from $\ket{\overline{1_t,0_{r}}}$ and $\ket{\overline{0_t,0_{r}}}$, respectively. As expected, the resonance conditions for this pair differ by the transmon’s anharmonicity.

Notably, our simulations reveal that the readout mode can influence the resonance condition of an intrinsic transmon transition. In particular, the $\ket{0_{t}}\leftrightarrow\ket{9_{t}}$ transition occurs in the transmon-only simulations at $\sim 11$ GHz, see Fig.~\ref{fig:landscape_floquet_g_e}(e) and Tab.~\ref{tab:inelastic_transitions}. However, as described in the previous section, the state $\ket{9_t, 0_r}$ strongly hybridizes with $\ket{6_t,1_r}$ in the joint system. As a result, $\ket{\overline{0_t, 0_r}}$ now experiences transitions to the hybridized states $\ket{\overline{21}}$ and $\ket{\overline{22}}$ instead. Complicating matters further, the resonance frequency of the $\ket{\overline{0_t, 0_r}} \leftrightarrow \ket{\overline{3_t, 1_r}}$ transition coincides with the above transitions (at zero drive amplitude); see Fig.~\ref{fig:app_inelastic_with_readout}(a,e,f). Thus, there is an interplay between all three transitions, which contributes to the observed transition line shapes in Fig.~\ref{fig:app_inelastic_with_readout}(a).

%%=============================================================
%% Appendix F: Identifying parasitic electromagnetic modes
%%=============================================================
\section{Identifying parasitic electromagnetic modes}
\label{sec:extrinsic_hfss}
%%-------------------------------------------------------------
%% Figure A5: Spurious packaging modes
%%-------------------------------------------------------------
\begin{figure}[t!]
\includegraphics[width=0.48\textwidth]{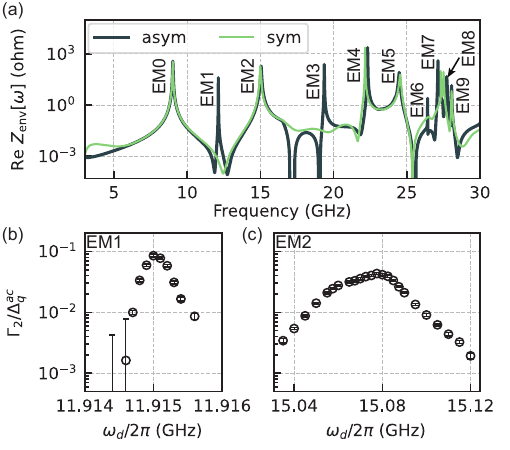}
\caption{Spurious packaging modes
(a) Environmental impedance from electromagnetic simulation of the package, assuming perfect symmetry (green) and with asymmetry along the long axis of the cavity (dark teal). Breaking of the geometric symmetry can cause the transmon couple to more packaging modes. 
(b-c) Spectroscopic evidence of the spurious geometric modes EM1 and EM2 identified in experiment.
}
\label{figA:geometric_modes} 
\end{figure}
%%-------------------------------------------------------------

In an experimental device, the circuit’s electromagnetic field is distributed in the three-dimensional space, inevitably giving rise to parasitic electromagnetic modes, which participate in the Josephson junction. 
As explained in Sec.~\ref{sec:geometric_inelastic}, these modes mediate inelastic scattering processes when certain resonance conditions are met by the drive. 

In principle, the spectrum of these parasitic modes can be obtained through electromagnetic simulations of the device geometry using finite-element solvers such as Ansys HFSS. In our setup, when the transmon is positioned precisely at the center of the 3D cavity, certain cavity modes remain uncoupled from it due to the symmetry of the design.  
This behavior is reflected in a \textit{driven modal simulation} performed in HFSS under ideal symmetric conditions.
The resulting environmental impedance, $\mathrm{Re}Z_{\mathrm{env}}(\omega)$, as seen by the transmon, is shown in Fig.~\ref{figA:geometric_modes}(a) as the light green trace. Notice that only a handful of peaks appear in this trace. 
The trace also indicates the absence of coupling to any modes between $9$ to $15$ GHz and  between $15$ to $22$ GHz in this case. 

However, in practice, this symmetry is often broken, either by misalignment of the qubit chip within the cavity, or by distortion of the electric fields caused by the introduction of coupling pins.
These asymmetries cause the transmon to couple to additional modes. When such imperfections are incorporated into the simulation, the resulting $\mathrm{Re}Z_{\mathrm{env}}(\omega)$ (dark teal trace in Fig.~\ref{figA:geometric_modes}(a)) displays several additional peaks. 
The modes labeled EM1 and EM3 in the simulation are identified from HFSS eigenmode simulation as the TE102 and TE202 cavity modes, respectively. In our device, they indeed couple to the transmon and mediate DUST under appropriate drive conditions. (See Table.~\ref{tab:cavity_modes}). 

%%-------------------------------------------------------------
%% Table A3: Spurious modes
%%-------------------------------------------------------------
\begin{table}[b]
    \centering
    \begin{tabular}{|c|c|c|}
        \hline 
        Transition(s) & Identified process(es) & Associated mode\\
        \hline 
        A & $1_t,0_{r} \rightarrow 0_t,1_{r}$ & EM2 \\
        \hline 
        B (B') & $1_t,0_{r} \rightarrow 2_t,1_{r}$ ($0_t,0_{r} \rightarrow 1_t,1_{r}$) & TLS \\
        \hline
        D (D') & $1_t,0_{r} \rightarrow 2_t,1_{r}$ ($0_t,0_{r} \rightarrow 1_t,1_{r}$) & RO $\equiv$ EM0 \\
        \hline
        G & $1_t,0_{r} \rightarrow 0_t,1_{r}$ & EM4 \\
        \hline 
        I (I') & $1_t,0_{r} \rightarrow 2_t,1_{r}$ ($0_t,0_{r} \rightarrow 1_t,1_{r}$) & EM2 \\
        \hline
        J & $1_t,0_{r} \rightarrow 0_t,1_{r}$ & EM5 \\
        \hline 
        K (K') & $1_t,0_{r} \rightarrow 8_t,0_{r}$ ($0_t,0_{r} \rightarrow 22$) & RO \\
        \hline
        L (L') & $1_t,0_{r} \rightarrow 4_t,1_{r}$ ($0_t,0_{r} \rightarrow 3_t,1_{r}$) & RO \\
        \hline
        N & $1_t,0_{r} \rightarrow 0_t,1_{r}$ & EM9 \\
        \hline
    \end{tabular}
    \caption{The correspondence between the observed transitions and the associated spurious modes inferred from the frequency-matching conditions for nonlinear wave-mixing processes. Note that the transition B(B') is not accounted for by any simulated electromagnetic modes. As elaborated in Appendix \ref{app:tls_mpt_dynamics}, they are mediated by a TLS. }
    \label{tab:cavity_modes}
\end{table}
%%-------------------------------------------------------------

To independently verify the presence of these modes and their coupling to the transmon, we perform spectroscopy measurements around the predicted frequencies of these spurious electromagnetic modes. 
We leverage the sensitivity of the transmon’s coherence times to photon shot noise in spurious RF modes. These modes couple dispersively to the qubit and when driven near resonance, the photon shot noise in these modes induces qubit dephasing.
Specifically, we extract the ac-Stark shift $\Delta_q^{\mathrm{ac}}$ and the induced dephasing rate $\Gamma_2$ by applying a weak probe tone with variable frequency and power during a Ramsey experiment. The probe is turned on continuously during the wait time between the two Ramsey pulses.

We extract the derivative $d\Gamma_2/d\Delta_q^{\mathrm{ac}}$ as a function of probe frequency.
This quantity directly reflects the environmental density of states and, equivalently, the real part of the environmental impedance $\mathrm{Re}Z_{\mathrm{env}}[\omega]$~\cite{Gambetta2006_dephasing}.
The result of this experiment, shown in Fig.~\ref{figA:geometric_modes}(b-c), exhibits resonant features at 11.915 GHz and 15.08 GHz. 
These experimentally observed resonances agree with the simulated EM1 and EM2 modes, respectively, confirming that these geometric modes couple to the transmon and can mediate drive-induced transitions.

Furthermore, assuming that the spectroscopic probe tone is in the weak-power limit ($\chi \sqrt{\bar{n}} \ll \kappa$), we can approximate $\Gamma_2 = 2\chi \Delta_q^{ac}/\kappa$~\citep{Schuster2005_dephasing,Gambetta2006_dephasing}. 
Using this relation, we estimate a dispersive coupling strength of $\chi/2\pi \approx 20$~kHz between EM1 and the transmon, and $\chi/2\pi \approx 0.4$~MHz between EM2 and the transmon. These values are essential input parameters for modeling the system using multi-mode Floquet simulations that include spurious electromagnetic resonances.

Note that although we identify EM2 as the TE103 cavity mode, its experimental line shape deviates from an ideal Lorentzian, as shown in Fig.~\ref{figA:geometric_modes}(c). This deviation suggests that EM2 is hybridized with an additional, unmodeled degree of freedom, possibly a standing wave mode in a mismatched cable or a structural resonance of the device package that is not captured in the finite-element model.

Finally, we identify and label all the transitions observed in the two-tone spectroscopy data that originate from mechanism \textbf{C}—inelastic scattering processes involving spurious electromagnetic modes. Each transition is associated with a specific mode responsible for mediating the process, and the full list is summarized in Table~\ref{tab:cavity_modes}.
For instance, mode EM2 is responsible for the transitions labeled \textbf{A} , \textbf{I} and \textbf{I\ensuremath{'}} in Fig.~\ref{fig:landscape_floquet_g_e}(a-b). 

Transition \textbf{A} originates from a \textit{4-wave-mixing, two-mode conversion} process, i.e. it is a $\ket{1_t, 0_{\rm{EM2}}} \rightarrow \ket{0_t, 1_{\rm{EM2}}}$ transition by the absorption of two drive photons. In contrast, transitions \textbf{I } and \textbf{I\ensuremath{'}} result from a pair of \textit{4-wave-mixing, two-mode squeezing} processes, namely, $\ket{1_t, 0_{\rm{EM2}}} \rightarrow \ket{2_t, 1_{\rm{EM2}}}$  and $\ket{0_t, 0_{\rm{EM2}}} \rightarrow \ket{1_t, 1_{\rm{EM2}}}$, accompanied by the absorption of two drive photons. 

As discussed in Appendix~\ref{app:floquet_extrinsic_details}, 
these spurious modes can also mediate higher-order mixing processes beyond four-wave mixing. For example, transitions \textbf{K}(\textbf{K\ensuremath{'}}) and \textbf{L}(\textbf{L\ensuremath{'}}) correspond to multi-photon processes that excite the transmon to higher levels such as $\ket{3_t}, \ket{4_t}$ or even $\ket{7_t}$.

%%=============================================================
%% Appendix G: Time dynamics of resonant exchange with a TLS
%%=============================================================
%%-------------------------------------------------------------
%% Figure A6: Dynamics of resonant exchange of the transmon 
%% excitation with the TLS environment.
%%-------------------------------------------------------------
\begin{figure}[t]
\includegraphics[width=0.48\textwidth]{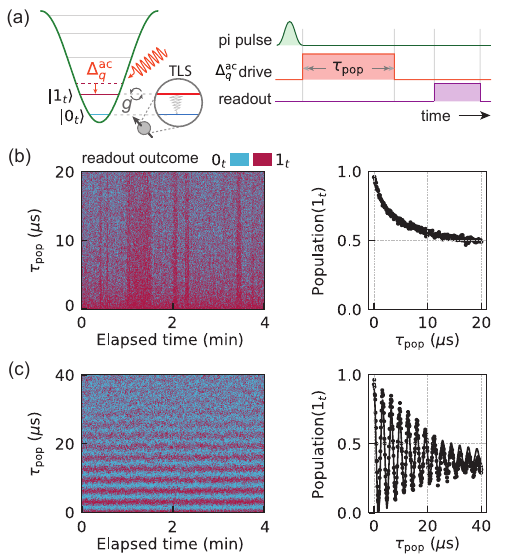}
\caption{Dynamics of resonant exchange of the transmon excitation with the TLS environment. (a) Experimental sequence, an ac Stark shift tone tuning the transmon on resonance with a TLS is played for a variable duration. 
(b) Time dynamics of resonant exchange with a strongly-dissipative $(g\ll\Gamma_1)$ TLS, resulting in an exponential decay of the transmon population. 
(c) Time dynamics of resonant exchange with a weakly-dissipative $(g \gg \Gamma_1)$ TLS, causing a decaying oscillation of the transmon population. 
}
\label{figA:TLS_exchange_dynamics} 
\end{figure}
%%-------------------------------------------------------------

\section{Time dynamics of resonant exchange with a TLS}
\label{sec:tls_decay_dynamics} 
As discussed in Sec. \ref{sec:tls_ac_stark_shift}, the static transmon-TLS interaction can be modeled by a linear XX coupling in the form of 
$g(\hat{b}^{\dagger} + \hat{b})\hat{\sigma}_x$, where $\hat{b}$ ($\hat{b}^{\dagger}$) is the annihilation (creation) operator of the transmon and $\hat{\sigma}_x$ is the Pauli X operation of the TLS, and $g$ is the coupling strength. Under the ac Stark shift, when the transmon tunes into resonance with the TLS, it can exchange energy with the TLS. Depending on the lifetime of the TLS, compared to the XX coupling rate, the population in the transmon will either show an oscillatory behavior or an exponential decay.
To gain further insight into the resulting dynamics of this interaction, we choose the power of the AC Stark shift tone to park the qubit on resonance with one particular TLS at a time. 
We then sweep the duration of this AC Stark shift tone, and measure the final transmon population as a function of this duration, as illustrated in Fig.~\ref{figA:TLS_exchange_dynamics}(a). From the time dynamics of the transmon population, we identify the two distinct classes of TLS, distinguished by the relative magnitude of the coupling strength $g$, and the TLS dissipation rate $\Gamma_1$. 

The resonant exchange with a strongly dissipative TLS $g \ll \Gamma_1$ will cause the transmon $1_t$ population to decay exponentially over time. 
An example is shown in Fig.~\ref{figA:TLS_exchange_dynamics}(b), with the transmon parked on resonance with a strongly coupled TLS at $\Delta_q^{ac}/2\pi = 135$ MHz. 
This is the same TLS leading to the feature at $|\Delta_q^{ac}/\alpha_q| = 0.73$ in main text Fig.~\ref{fig:tls_spectrum}(c). 
The enhanced decay is present when the TLS switches into resonance with the transmon, and absent when the TLS switches away. 
Such frequency-switching behavior is evident in the unaveraged data monitored over $4$ minutes, shown in the left panel. 
The background for the average population decay shown in the right panel is determined by the thermal population of the TLS as well as the on-off ratio for this resonant exchange. 

In contrast, for a TLS with $g \gg \Gamma_1$, the transmon population exhibits coherent oscillations, indicating the coherent exchange of quanta between the transmon and TLS on the timescale $\sim 1/g$. 
An example is shown in Fig.~\ref{figA:TLS_exchange_dynamics}(c), with the transmon parked at $\Delta_q^{ac}/2\pi =  18$ MHz, on resonance with the TLS associated with transition \textbf{P} in Fig.~\ref{fig:landscape}(c).
From the oscillating pattern, we extract a $\Gamma_1 = 100\mu \rm{s}^{-1}$ for this TLS, with a coupling rate $g = 1.6\mu \rm{s}^{-1}$ to the transmon. 
Note that this TLS disappeared when the data for main text Fig.~\ref{fig:tls_spectrum}(c) was being taken. 

%%=============================================================
%% Appendix H: Time dynamics of an intrinsic 
%% multi-photon excitation
%%=============================================================
\section{Time dynamics of an intrinsic multi-photon excitation}
\label{sec:time_dynamics_intrinsic}
%%-------------------------------------------------------------
%% Figure A7: Dynamics of an intrinsic multi-photon excitation 
%%-------------------------------------------------------------
\begin{figure}[t]
\includegraphics{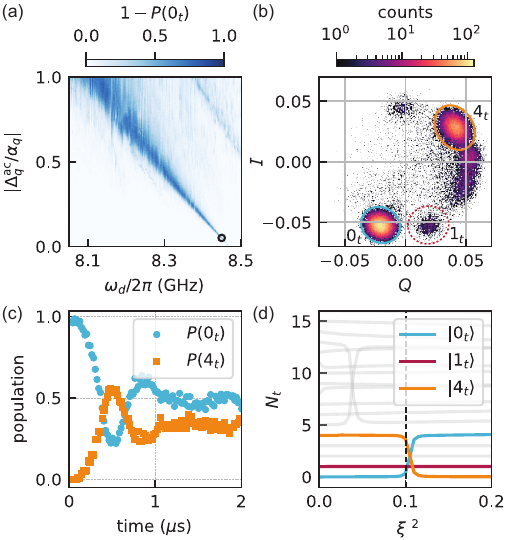}
    \caption{Dynamics of an intrinsic multi-photon excitation process. (a) Zoom in of the feature \textbf{E\ensuremath{'}} from the landscape Fig.~\ref{fig:landscape_floquet_g_e} (b) Readout histogram of all the shots from the time domain experiment, showing a dominant population transfer into $\ket{4_t}$.  (c) Temporal dynamics showing oscillations of the final population between $|0_t\rangle$ and $|4_t\rangle$ as a function of $\tau_{\rm{pop}}$ when driven on resonance. (d) Floquet branch analysis indicating the hybridization between $\ket{\tilde0_t}$ and $\ket{\tilde4_t}$ at drive frequency $\omega_d/2\pi$ = 8.45 GHz.}
\label{figA:04transition_dynamics}
\end{figure}
%%-------------------------------------------------------------

We investigate the dynamics of the intrinsic multi-photon transitions by performing a time-domain experiment.
As an example, we illustrate the feature labeled \textbf{E\ensuremath{'}} in Fig.~\ref{fig:landscape_floquet_g_e}. This feature corresponds to a transition where two drive photons are absorbed to excite the transmon $|0_t\rangle \rightarrow |4_t\rangle$. 
We apply the same pulse sequence as Fig.~\ref{figA:TLS_exchange_dynamics}(a), parking the frequency and the power of the stimulation tone on resonance with feature \textbf{E\ensuremath{'}}. 
We choose the drive frequency of $\omega_d/2\pi$ = 8.45 GHz, for which the resonance is activated at a small drive power, $\xi^2 \simeq0.1$, thus avoiding any hybridization with other higher levels.
The drive condition for this time-domain experiment is marked by a circle in Fig.\ref{figA:04transition_dynamics}(a). 
When we perform a readout following this multi-photon resonance drive, we observe a direct population transfer to $\ket{4_t}$ of the transmon, without significant population in the intermediate levels, as shown by the readout histogram, plotted in Fig.~\ref{figA:04transition_dynamics}(b). The small residual populations in $\ket{3_t}$ and $\ket{2_t}$ arise from incoherent decay processes from  $\ket{4_t}$ during the experiment. 

%%-------------------------------------------------------------
%% ADJUSTED FOR PLACEMENT
%% Figure A8: Single-quantum transmon excitation involving an 
%% extrinsic mode with a switching frequency
%%-------------------------------------------------------------
\begin{figure}[t]
\includegraphics{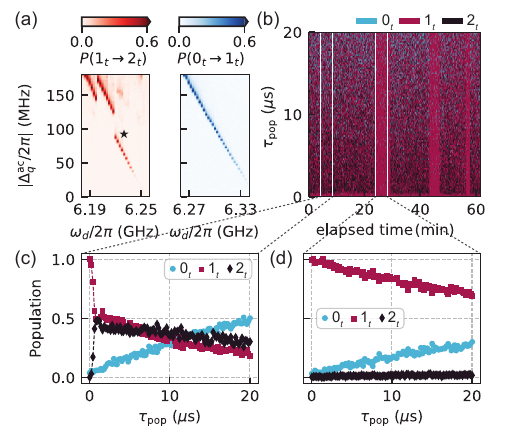}
\caption[Multi-quanta resonance involving a switching TLS]{Single-quantum transmon excitation involving an extrinsic mode with a switching frequency. (a) Probability of transmon excitation. The features \textbf{B} and \textbf{B\ensuremath{'}} are caused by the same extrinsic mode. (b) Single-shot readout outcome data string showing $\{0_t, 1_t, 2_t\}$, taken at the drive condition marked star in (a). The states are identified with a precalibrated IQ multithreshold. (c-d) Average time dynamics within two separate time windows, indicated by the vertical white lines in (b). The excitation to $2_t$ is present when the mode switches into the resonance (shown in c), and absent when the mode switches away (shown in d).
}
\label{figA:TLS_heating_2} 
\end{figure}
%%-------------------------------------------------------------
To obtain the temporal dynamics, we first define readout thresholds in the quadrature plane, defined by the  $3\sigma$ boundaries for the readout outcomes, $\ket{0_t}$ and $\ket{4_t}$. These thresholds are shown by the blue and orange ellipses with solid boundaries.
By turning on the drive for a variable duration $\tau_{\rm{pop}}$, we observe a coherent oscillation of the transmon population between $\ket{0_t}$ and $\ket{4_t}$, as plotted in Fig.~\ref{figA:04transition_dynamics}(c). 

Through this experiment, we can extract the rate of the multi-photon resonances, as a function of drive frequencies and powers, and also investigate the excited-state dynamics~\cite{Wang2025} during DUST. However, hybridization of this transition with other processes, and the resulting offset charge sensitivity limit one to low drive powers for such rate extraction experiments. This limitation can be circumvented by designing transmons with $E_J\ggg E_C$.

The Floquet simulation confirms this process by showing a \textit{branch swap} as illustrated in Fig.~\ref{figA:04transition_dynamics}(d). 
The observed dynamics can be understood as a manifestation of Landau-Zener-Stückelberg interference~\cite{SHEVCHENKO2010} between the two hybridized Floquet modes. 

%%-------------------------------------------------------------
%% Figure S9: Single-quantum transmon excitation involving an 
%% extrinsic mode with a switching frequency
%%-------------------------------------------------------------
\begin{figure*}[t]
\includegraphics[width=0.96\textwidth]{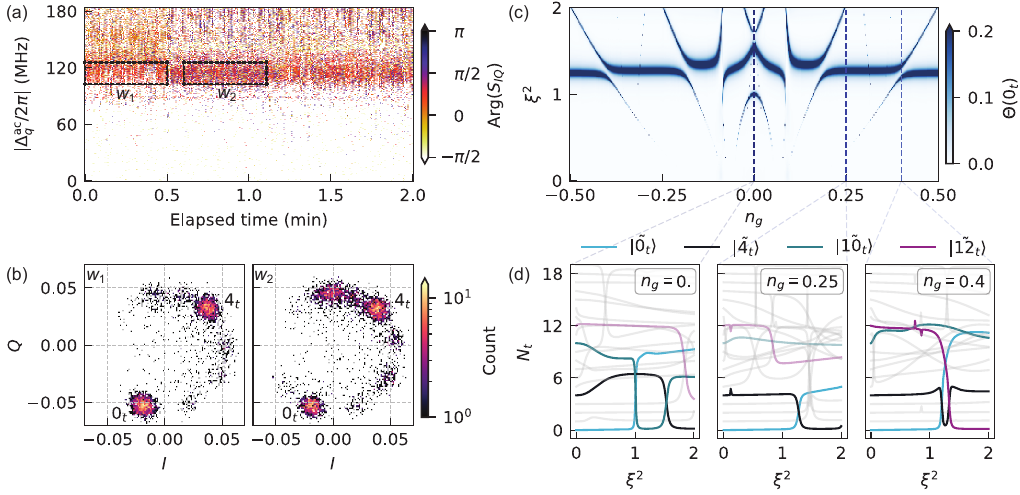}
\centering
\caption{Offset charge dependence of intrinsic multi-photon transitions. (a) Phase of the single-shot readout IQ signals as a function of drive power at $\omega_d/2\pi = 8.2325$ GHz. The transmon undergoes state transitions above drive powers that correspond to an ac Stark shift $|\Delta_q^{\rm ac}/2\pi|\approx110$ MHz. When tracked over a duration of $2$ minutes, the resulting readout signal shows different phases in the IQ plane, indicating different final states.
(b) Histogram of the readout signals sampled from two different windows $w1$ and $w2$ in (a), confirming temporally drifting final states after the transition. Even though the dominant transition happens to $\ket{4_t}$, multiple higher states get involved and their contribution fluctuates in time. (c) The hybridization parameter $\Theta(0_t)$ from the Floquet simulation as a function of offset charge on the island and the drive strength, showing offset charge dependence of the intrinsic multi-photon transition. (d) Branch analysis performed at three different  offset charge conditions, $n_g = \{0, 0.025, 0.4\}$, shown by the vertical dashed lines in (c). In addition to the branch swap between $\ket{\tilde{0}_t}$ and $\ket{\tilde{4}_t}$, at different offset charge conditions, additional branch swaps involve the levels $\ket{\tilde{10}_t}$ and $\ket{\tilde{12}_t}$, confirming the experimental observation.
}
\label{figA:ng_drift} 
\end{figure*}
%%-------------------------------------------------------------

%%=============================================================
%% Appendix I: Time dynamics of the inelastic
%% scattering mediated by a TLS
%%=============================================================
\section{Time dynamics of the inelastic scattering mediated by a TLS}
\label{app:tls_mpt_dynamics}

We identify a pair of single-quantum excitation features, i.e., $\ket{0_t}\rightarrow\ket{1_t}$ and $\ket{1_t}\rightarrow\ket{2_t}$, labeled \textbf{B\ensuremath{'}} and \textbf{B}, respectively, in Fig.~\ref{fig:landscape_floquet_g_e}(a-b) that are not explained by any of the electromagnetic modes we simulated up to $30$ GHz. We zoom into these two features and replot it in  Fig.~\ref{figA:TLS_heating_2}(a). We already discussed in Sec.~\ref{sec:inelastic_tls} that feature \textit{B} in due to an inelastic scattering process that involves a switching TLS at $~8.17$ GHz. \textbf{B\ensuremath{'}} is the corresponding transition involving the same TLS when the transmon in initialized in $\ket{0_t}$.

To confirm this interpretation, we examine the time dynamics of the transmon populations under the drive condition marked by the asterisk in Fig.~\ref{figA:TLS_heating_2}(a). 
Unaveraged, single-shot measurements reveal a clear telegraphic ``on-off'' switching behavior of the $\ket{1_t} \rightarrow \ket{2_t}$ transition, as shown in Fig.~\ref{figA:TLS_heating_2}(b). 
We plot the averaged time dynamics during two different windows, defined by the  white vertical lines, and observe qualitatively distinct behaviors.
When the TLS is in the higher-frequency state, satisfying the resonance condition for this process,
the qubit is rapidly excited to $\ket{2_t}$ [Fig. \ref{figA:TLS_heating_2}(c)]; 
When the TLS switches to the lower-frequency state, the drive is off from the resonance condition, and the transmon remains in $\ket{1_t}$, and is only subjected to a slower natural $T_1$ decay to $\ket{0_t}$, [Fig.~\ref{figA:TLS_heating_2}(d)]. 
The timescale of switching of this TLS is several minutes, similar to that of the slow switching TLS we observe near the qubit frequency, as shown in Fig.~\ref{fig:tls_spectrum}.

Note that, The $\ket{0_t} \rightarrow \ket{1_t}$ data shown here was acquired during a different time window and the TLS did not switch during this period. 

%%=============================================================
%% Appendix J: Ng sensitivity of intrinsic multiphoton 
%% transitions
%%=============================================================

\section{Offset charge sensitivity of intrinsic multi-photon transitions}
\label{app:ng_drift}
Transitions arising from intrinsic multi-photon excitations (such as E, F, and M) in the spectroscopy landscape appear significantly broadened and ``fuzzy'' at higher drive power, as seen in Fig.~\ref{fig:landscape_floquet_g_e}. 
Such behavior is due to the temporal fluctuations of the multi-photon resonance conditions due to the $n_g$ dependence of such transitions. As an example, we revisit the transition labeled \textbf{E}' in Fig.~\ref{fig:landscape_floquet_g_e}, which corresponds to a process that absorbs two drive photons to excite the transmon from $|0_t\rangle$ to $|4_t\rangle$.

In Fig.~\ref{figA:ng_drift}(a) we show the unaveraged single-shot measurement results from a two-tone spectroscopy experiment at a fixed drive frequency of 8.2375 GHz, monitored over the course of 2 minutes. We plot the phase of the integrated IQ signal as a function of drive power and elapsed time. We observe that the resonance condition itself drifts in drive power, leading to a 
fuzzy transition, distributed over a range of drive powers that corresponds to an ac Stark shift between $100$ and $180$ MHz.
Such fluctuations occur on a sub-ms time scale. Moreover, when such a transition happens, the readout phase has different values at different times, as illustrated in two different time windows, $w_1$ and $w_2$, for the same range of drive powers. This distinct change indicates that the transition leads to different final states at different times. We confirm this observation by comparing the readout histograms constructed from these two different time windows, as shown in Fig.~\ref{figA:ng_drift}(b). Although the transition dominantly excites the transmon to $\ket{4_t}$, multiple higher excited states also get populated for this drive condition, and the probability of exciting the transmon to this higher excited state drifts over time.

Even though the offset charge sensitivity of the static transmon level $\ket{4_t}$ is expected to be less than a few hundred kHz, the driven transmon level $\ket{\tilde{4}_t}$ becomes significantly offset charge sensitive. This increased sensitivity arises from the drive-induced hybridization of the transmon levels. As the offset charge fluctuates between experiments, the resonance condition for the intrinsic multi-photon transition, $\ket{0_t}\rightarrow\ket{4_t}$, also fluctuates, giving rise to the observed fuzzy resonance in Fig.~\ref{figA:ng_drift}(a).

To support our interpretation of these results, we perform Floquet simulation for the same drive frequency 8.2375 GHz with sweeping $n_g$. 
In Fig.~\ref{figA:ng_drift}(c), we show the hybridization parameter for the ground state $\Theta(0_t)$, plotted for the range of drive power that corresponds to the same range of ac Stark shift in the experiment. The peaks in the hybridization parameter $\Theta(0_t)$ correspond to multi-photon resonances from $\ket{\tilde{0}_t}$.
We observe that the resonance condition varies significantly with offset charge $n_g$, spanning a wide range in drive power. 
Importantly, it exhibits avoided crossings as $n_g$ is swept, indicating that the shift in the observed resonance results from hybridization with higher-lying levels which are $n_g$ sensitive. When the drive condition satisfies such a transition to several hybridized levels, the transmon ends up in a mixture of several higher-energy states.

To identify these final states, we perform Floquet branch analysis at different offset charge conditions. We show the corresponding plots for three representative offset charges, $n_g = \{0, 0.25, 0.4\}$ in Fig.~\ref{figA:ng_drift}(d). In the branch analysis plot for offset charge $n_g=0$ and $n_g=0.4$, we observe that level $\ket{\tilde{0}_t}$ undergoes branch swaps not only with $\ket{\tilde{4}_t}$, but also with $\ket{\tilde{10}_t}$ and $\ket{\tilde{12}_t}$ respectively.
The sensitivity of these resonances to offset charge thus provides a microscopic explanation for the temporal fluctuations in the observed transitions and the measured population in highly-excited states in the experiment.

%%=============================================================
%% BIBLIOGRAPHY
%%=============================================================
%apsrev4-2.bst 2019-01-14 (MD) hand-edited version of apsrev4-1.bst
%Control: key (0)
%Control: author (8) initials jnrlst
%Control: editor formatted (1) identically to author
%Control: production of article title (0) allowed
%Control: page (0) single
%Control: year (1) truncated
%Control: production of eprint (0) enabled
%

\end{document}